\documentclass[lettersize,journal]{IEEEtran}
\usepackage{amsmath,amsfonts}
\usepackage{algorithmic}
\usepackage{algorithm}
\usepackage{array}
\usepackage[caption=false,font=normalsize,labelfont=sf,textfont=sf]{subfig}
\usepackage{textcomp}
\usepackage{stfloats}
\usepackage{url}
\usepackage{verbatim}
\usepackage{graphicx}
\usepackage{pifont}
\usepackage{multirow}
\usepackage{booktabs}
\usepackage{arydshln} % 导入绘制虚线的宏包

\usepackage{cite}
\usepackage{siunitx}
\usepackage{booktabs}
\usepackage[colorlinks = true,
            linkcolor = blue,
            urlcolor  = blue,
            citecolor =  green,
            anchorcolor = blue]{hyperref}
\usepackage{xcolor}

\begin{document}

\newcommand{\vModelName}{\textit{AudioLDM 2}}

\title{DreamAudio: Customized Text-to-Audio Generation with Diffusion Models}

\author{Yi Yuan, Xubo Liu, Haohe Liu~\IEEEmembership{Graduate Student Member,~IEEE}, Xiyuan Kang, Zhuo Chen\\  Yuxuan Wang, Mark D. Plumbley,~\IEEEmembership{Fellow,~IEEE},  Wenwu Wang,~\IEEEmembership{Fellow,~IEEE}  \\ 
% \url{https://audio.github.io/audio2/}
\thanks{Yi Yuan, Xubo Liu, Haohe Liu, Xiyuan Kang, and Wenwu Wang are with the School of Computer Science and Electronic Engineering, University of Surrey, Guildford, UK. Email: \{yi.yuan, xubo.liu, haohe.liu, xk00063, m.plumbley, w.wang\}@surrey.ac.uk.}
\thanks{ Mark D. Plumbley is with Department of Informatics, King's College London, London, UK. Email: mark.plumbley@kcl.ac.uk.}
\thanks{Zhuo Chen and Yuxuan Wang are with the Seed Group, ByteDance Inc. Email: \{zhuo.chen1, wangyuping, wangyuxuan.11\}@bytedance.com.}
\thanks{Codes and demos are available at~\url{https://yyua8222.github.io/DreamAudio_demopage/}.}
}


\maketitle

 \begin{abstract}
With the development of large-scale diffusion-based and language-modeling-based generative models, impressive progress has been achieved in text-to-audio generation. Despite producing high-quality outputs, existing text-to-audio models mainly aim to generate semantically aligned sound and fall short of controlling fine-grained acoustic characteristics of specific sounds. As a result, users who need specific sound content may find it difficult to generate the desired audio clips. In this paper, we present DreamAudio for customized text-to-audio generation~(CTTA). Specifically, we introduce a new framework that is designed to enable the model to identify auditory information from user-provided reference concepts for audio generation. Given a few reference audio samples containing personalized audio events, our system can generate new audio samples that include these specific events. In addition, two types of datasets are developed for training and testing the proposed systems. The experiments show that DreamAudio generates audio samples that are highly consistent with the customized audio features and aligned well with the input text prompts. Furthermore, DreamAudio offers comparable performance in general text-to-audio tasks. We also provide a human-involved dataset containing audio events from real-world CTTA cases as the benchmark for customized generation tasks.

\end{abstract}


\section{Introduction}

\IEEEPARstart{A}udio generation, as a crucial technology for enabling artificial intelligence generated content~(AIGC)~\cite{cao2023comprehensive}, has gained significant interest from the research community. In recent years, conditional audio generation has become a popular paradigm, where music, speech, or general sound effects are generated based on a variety of conditions, such as text~\cite{audioldm}, images~\cite{sung2023sound,sheffer2023hear-image-to-audio}, and videos~\cite{foleygen,specvqgan}. This opened up new opportunities for a range of potential applications, including audio generation for movies~\cite{kreuk2022audiogen}, games~\cite{riffusion}, and audiobooks~\cite{liu2023wavjourney}.

% , and followingground sounds for music productions~\cite{kreuk2022audiogen}.  

Facilitated by advances in diffusion-based generative models~\cite{ddim,DDPM,valle2025fugatto} and large-scale audio-language datasets~\cite{audioset,wavcaps,autoacd,sound_vecaps}, several models have been developed for audio generation with text prompts as conditions, such as AudioLDM~\cite{audioldm}, AudioGen~\cite{audiogen}, DiffSound~\cite{diffsound}, TANGO~\cite{tango}, Make-an-Audio2~\cite{makeanaudio2}, Re-AudioLDM~\cite{reaudioldm}, and AudioLDM2~\cite{audioldm2}. Building on the ``semantic prior'' learned from large collections of datasets~\cite{dreambooth} to associate textual concepts with audio features, these models have shown strong capabilities in generating audio samples of high quality, fidelity and diversity. For example, when provided with the text prompt \textit{dog barking}, the model leverages this particular ``semantic prior'' to generate various audio clips depicting \textit{dog barking} across varying species, emotions, and durations. %but fails to generate specific \textit{dog barking} based on user preferences.

Despite significant progress in diffusion-based methods~\cite{dalle,DALLE2,dalle3,stablediffusion,stable-diffusion-3}, current text-to-audio generation systems often lack the flexibility to customize content based on personalized intentions. This can cause problems in real-world multimedia production, which often requires the generation of audio samples to be tailored for specific features. For example, it can be challenging for current text-to-audio~(TTA) models to generate sounds that are consistent with the text prompt \textit{``a monster is fighting with a Minion''}. This is because audio events such as ``monster fight'' and ``Minion talk'' are unique and rare, or with a specific timbre, which can hardly be found in any current training data. Hence, users often need to go through multiple rounds of trial and error to produce the desired output and may encounter difficulties in achieving the optimal result~\cite{saito2024soundctm}.

\begin{figure}[t]
  \centering
  \includegraphics[width=0.49\textwidth]{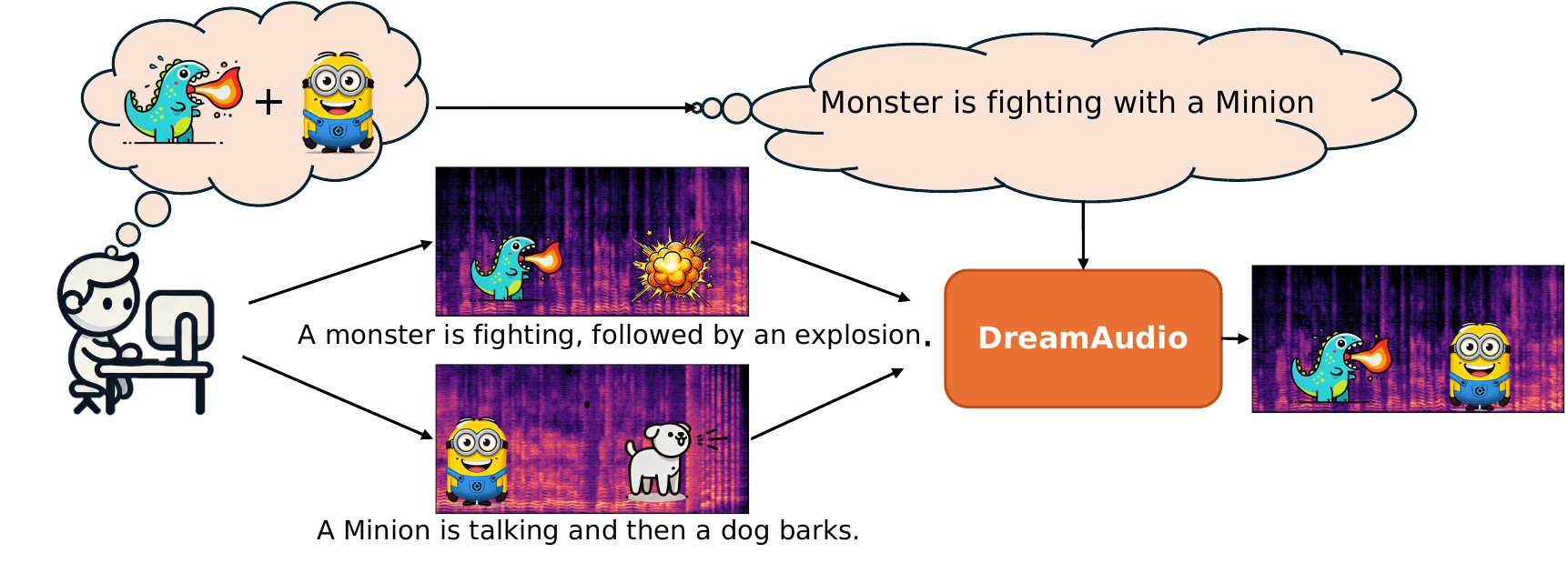}

  \caption{An illustration of \textit{DreamAudio} for audio generation with customized content of ``monster fighting'' and ``Minion talking''. The system takes both the text prompt and user-provided audio-caption pairs as the reference concepts, and generate audio content consistent with the description ``Monster is fighting with a Minion". }
  \label{fig:init_demo}
    \vspace{-15pt}
\end{figure}

To address this challenge, several works have explored the ideas for generating rare or unseen audio events, especially for few-shot or zero-shot scenarios. One of the pioneering works is the Re-AudioLDM~\cite{reaudioldm}, which applies retrieval-based techniques to improve the performance for the generation of rare audio events using audio-caption pairs retrieved from external data. Similarly, AudioBox TTA-RAG~\cite{audiobox-tta} introduced a retrieval embedding module to improve the performance for few-shot and zero-shot audio generation tasks. Although these previous approaches effectively improve the semantic accuracy in generating infrequent or unseen audio events, they are still unable to provide explicit control over the sound effects generated, as specified by users.  

In this paper, we propose a new TTA task, namely, customized text-to-audio generation~(CTTA), where the audio content produced by the generation system can be controlled and customized by users. For instance, the system is capable to generate the \textit{dog barking} sound from a specific dog with unique timbre, or the sound for a specific \textit{monster fighting} which is not included in the training dataset. We refer to these featured audio events as user reference concepts. %For instance, existing models may generate one desired audio event across many trials~(e.g., one out of hundreds of attempts), while CTTA aims to reliably produce audio samples that consistently match user-specified characteristics.
To achieve this, we introduce \textit{DreamAudio}, a latent diffusion-based system with flow matching for the CTTA task. Inspired by ControlNet~\cite{controlnet}, we propose a multi-reference customization~(MRC) structure for the generator module to identify the reference concepts and control the generated content. More specifically, we design a new group of encoder blocks to extract features from the user-provided reference concepts. The system is trained to fuse such customized features for the generation of the audio output. \textcolor{black}{By incorporating multiple cross-attention modules that link the target text prompt with the corresponding reference text, our proposed approach establishes a pipeline capable of generating user-preferred audio samples without requiring concept-specific fine-tuning during inference. In contrast to tuning-based customization methods, which necessitate model updates for each new reference even after the model has been fully trained, DreamAudio directly extracts and integrates features from the new references within a single forward pass.
} 
Figure~\ref{fig:init_demo} shows an example of CTTA, where the model is enabled to generate the audio sample with customized ``monster fighting'' and ``Minion talking'', in terms of the provided reference concepts. 
%Leveraging the advantage of rectified flow matching~(RFM) on generative frameworks, our goal is to guide the model to capture the audio feature from the reference concepts and integrate this information into the generated outputs. 

To facilitate model development and evaluation, we create two datasets with different formats by concatenating and overlapping different audio events, called Customized-Concatenation and Customized-Overlay. In addition, we collect several special audio events and then manually design various customized cases, creating a small-scale dataset that more closely reflects real-world multimedia production scenarios and serves as a benchmark for CTTA. Experiments conducted on these datasets demonstrate that our method effectively handles customized audio generation and establishes a strong baseline for this new CTTA task.

% Furthermore, \textit{DreamAudio} presents comparable results to previous SOTA models on general text-to-audio generation tasks using AudioCaps~\cite{audiocaps}, illustrating substantial advancements in customized TTA tasks and competitive performance in general TTA tasks.

Our contributions can be summarized as follows.

\begin{list}{\labelitemi}{\leftmargin=1em}
    \item We propose a novel audio generation model, DreamAudio, that is capable of performing content-customized audio generation with text prompts as condition. 
    \item We propose a multi-reference customization (MRC) structure, which allows the features from the reference audio to be fused with input prompts for text-to-audio generation. 
    \item We develop two new datasets for model training and evaluation, and establish a new benchmark for content-customized text-to-audio generation. 
    \item Our experiments show that \textit{DreamAudio} significantly enhances the ability of TTA models for customized audio generation and establishes a strong baseline for this new CTTA task.
\end{list}

\section{Related Work}

\subsection{Diffusion Models}

\noindent
Diffusion-based models~\cite{DDPM, SGM} have demonstrated improved performance in generative tasks for image~\cite{DiffusionBeatsGANs, DALLE2, Imagen,ISRIR}, audio~\cite{WaveGrad,DiffWave,liu2024audiolcm}, and video~\cite{MakeAVideo, ImagenVideo}. In the realm of audio synthesis, researchers initially followed the design used in image generation, and adapted it for audio generation based on mel-spectrogram~\cite{popov2021gradtts, ResGrad} and waveform~\cite{BDDM, PriorGrad, InferGrad}. Due to the involvement of high-dimensional data~\cite{kong2019acoustic,Liu-tts}, the representations for waveform and spectrogram can be difficult to train and slow to infer. To address this limitation, recent diffusion-based models work in a latent space through encoding pipelines~\cite{stablediffusion}. Specifically, current state-of-the-art (SOTA) systems follow an encoder-decoder framework~\cite{leveraging}, trained to generate the target features in a latent space, which are then decoded into waveforms through vocoders~\cite{hifigan,melgan,bigvgan}. 

\subsection{Conditional Audio Generation}

\noindent
Conditional generation has emerged as an important area in audio synthesis, enabling users to guide generative models with specific constraints or conditions to produce desired outcomes~\cite{controlnet}. Text-driven audio generation has gained significant attention.  AudioGen~\cite{audiogen} employs a conditional language modeling pipeline that generates audio waveforms directly. AudioLDM~\cite{audioldm} employs the Contrastive Language-Audio Pretraining~(CLAP)~\cite{clap} model to generate the embeddings of audio and text. These embeddings are then used as conditions to guide the training and inference of the latent diffusion model for audio generation. Specifically, AudioLDM is trained to generate feature representations of target audio within the latent space, followed by a variational autoencoder~(VAE) decoder to reconstruct the spectrogram from the latent representation.
Make-an-Audio~\cite{makeanaudio2} develops a pseudo-prompt enhancement strategy to generate extra audio-caption pairs with large-scale compositions to alleviate the data scarcity problem. Tango~\cite{tango} uses a structure similar to AudioLDM, but replaces the CLAP model with Flan-T5~\cite{t5}. 

Due to the diversity of the training dataset~\cite{audioset,audiocaps,wavcaps}, these previous methods allow the model to generate highly diverse samples for the same prompt. However, these models are not designed for CTTA tasks and often struggle to generate content in terms of user preferences~\cite{reaudioldm}, which is the problem to be addressed in this paper. 
%In this paper, we propose a new framework for controllable audio generation, which is designed to generate audio samples, customized via audio concepts provided by users.

\begin{figure*}[t]
  \centering
  \includegraphics[width=\textwidth]{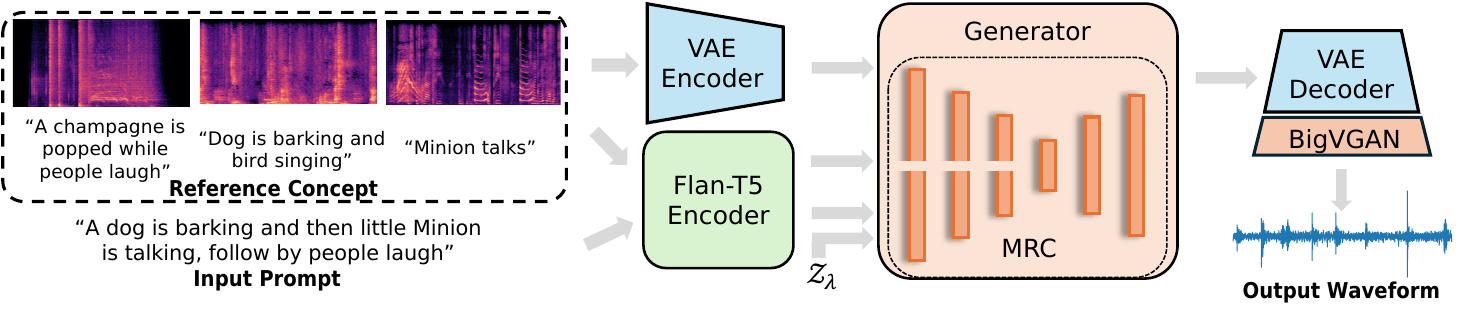}
  \caption{The inference pipeline of the \textit{DreamAudio}. The input prompt and reference concept are encoded in two paralleled paths through the Flan-T5 Encoder and the reference audio feature is encoded by the VAE Encoder. Along with the noisy data $\boldsymbol{z}_{\lambda}$, four inputs are forwarded to the generator with the MRC structure to generate the denoised data, followed by the VAE decoder and vocoder to reconstruct the final output waveform.}
  \label{fig:overview}
\end{figure*}

\subsection{Audio Generation with Flow Matching}

Flow matching has recently emerged as an efficient alternative to diffusion-based generative models. By learning a deterministic flow path between noise and target distribution~\cite{flowsep}, flow matching significantly reduces the number of inference steps required for high-quality synthesis, leading to improved efficiency and enhanced performance. FlashAudio~\cite{flashaudio} is the first audio generation model to employ flow matching, providing fast and high-fidelity results. LAFMA~\cite{lafma} further exploits flow-matching models and reduces the number of inference steps to ten without compromising performance. More recently, TangoFlux~\cite{tangoflux} and Stable Audio~\cite{stableaudio} have applied flow matching to large-scale text-to-audio models, achieving SOTA performance on several audio-generation benchmarks. These studies demonstrate that flow matching provides an efficient and scalable framework for audio synthesis, enabling faster sampling and strong performance. Our work builds upon this framework and applies flow matching as the core backbone.
\subsection{Customized Generation}
% \subsubsection{Image Generation}
Current methods for customized content generation are developed dominantly for images\cite{pre-training1, foley, masactrl,anydoor,encoder-based,imagic,key_locked}. Customized image generation could be categorized into tuning-based customization~\cite{dreambooth,inversion,XTI,NETI}, and tuning-free customization~\cite{cones,hyperdreambooth,elite,masactrl,freecustom,choi2023custom}. In tuning-based methods, such as DreamBooth~\cite{dreambooth}, the generation models are fine-tuned with the embeddings of specific subjects to control the generated contents. In tuning-free methods, such as FreeCustom~\cite{freecustom}, customization is achieved by integrating the target feature with the generated feature during the inference stages to minimize the fine-tuning processes. \textcolor{black}{Our DreamAudio resembles tuning-free customization methods. While the model itself is trained from scratch to learn general customization capabilities, it does not require any additional adaptation or parameter updates when encountering unseen user-provided reference concepts during inference.}

\subsubsection{Speech and Music Generation}
Several customized systems have been explored in speech and music generation. For speech generation, ViT-TTS~\cite{vit-tts} introduces a visual-text encoder that extracts additional visual scene information from reference images to improve text to speech (TTS) generation performance. F5-TTS~\cite{f5-tts} incorporates reference speech as conditioning signals to control speaking style, prosody, and voice characteristics. In the music domain, Plitsis et al.~\cite{plitsis2024investigating} adapted user-specific concepts from DreamBooth~\cite{dreambooth} and fine-tune the text embeddings for personalized music style control.
Unlike CTTA systems, these existing methods mainly adjust global acoustic attributes such as speaker style, prosody, timbre, genre, or overall musical texture. Their conditioning signals depend on complete speech or full musical phrases, whereas CTTA requires fine-grained, event-level controllability, making it fundamentally different from these style-based generation tasks.

\subsubsection{General Audio Generation} FreeAudio~\cite{freeaudio} and TG-Diff~\cite{evans2024fast} are two works designed to customize the timing of sound events in general audio generation. These two models both applied tuning-free customization strategies. More specifically, TG-Diff achieves temporal controllability by learning per-second embeddings, while FreeAudio achieves customization by applying a strategy for decoupling and aggregating attention tokens to fill-in the time-window tokens. However, both systems focus on timing control in text-to-audio generation, which differs from our goal on controlling the acoustic audio events. In addition, these two methods focus on training the time-related embeddings and tokens to match the timing window. Such requirement does not involve zero-shot scenarios, as most temporal patterns can be covered by massive training data with different temporal features.

\subsection{Retrieval-Based Generation}
Another task similar to CTTA is retrieval-based TTA, which was developed recently to improve the performance of TTA models in the few-shot and zero-shot cases. An example is Re-AudioLDM~\cite{reaudioldm}, which incorporates external audio features into the generation process using retrieval-augmented approaches~\cite{sheynin2022,reimagen}. This system demonstrates more stable performance on low-occurrence audio events in TTA generation. Building on this concept, AudioBox TTA-RAG~\cite{audiobox-tta} introduces a retrieval-information embedding module to enhance the capability in zero-shot generation. Although these systems can improve the TTA generation performance in the few-shot and zero-shot cases, the quality of the audio samples is limited by the database for audio retrieval. Furthermore, these models lack the capability to generate specific content for customization tasks, and research for CTTA tasks remains to be explored. 

In this paper, we design a novel CTTA system, which allows the generated audio content to be tailored and controlled in terms of user preference derived from the reference concepts. The ability to achieve customization not only enhances the creative process, but also improves controllability in the event level in TTA systems. Hence, we aim to address a crucial practical challenge for deploying TTA systems in real-world applications.

\section{Proposed Method} 
\label{sec: DreamAudio}
 % This section first defines the CTTA task in detail, then presents the approaches of the proposed model， and finally introduces several training strategies designed to enhance the performance of the system.

\subsection{Overview of the Proposed Model}
\label{sec: overview}
Given only a set of $k=1,...,K$ audio-language reference pairs $\{(\mathbf{a}_1, \mathbf{t}_1), (\mathbf{a}_2, \mathbf{t}_2), \dots, (\mathbf{a}_K, \mathbf{t}_K)\}$, where each reference audio clip $\mathbf{a}_k$ is associated with a reference caption $\mathbf{t}_k$, the CTTA task aims to generate the target audio $\mathbf{a}$ conditioned on the input textual prompt $\boldsymbol{c}$ and the reference concepts, derived from pairs of reference audios $\{\mathbf{a}_1, ..., \mathbf{a}_K\}$ with their corresponding captions $\{\mathbf{t}_1, ..., \mathbf{t}_K\}$.

To address this task, we propose the DreamAudio system, as illustrated in Figure \ref{fig:overview}. The system is composed of several modules, including the text encoder (e.g. Flan-T5~\cite{t5}) to obtain the embeddings of the input text prompt and reference captions, and the audio encoder (e.g. the variational autoencoder (VAE) as built in~\cite{audioldm}) to obtain the embeddings of the reference audios, which are followed by the feature generator built on rectified flow matching (RFM)~\cite{liuflow}. To customize the audio features with reference concepts, a multi-reference customization (MRC) structure is designed for the feature generator. The customized audio embeddings are then converted to spectrograms using a VAE decoder, which are then turned into waveforms using a vocoder (e.g. BigVGAN~\cite{bigvgan}).    

In the remainder of this section, we first discuss the calculation of text and audio embeddings. Then, we present our framework for diffusion-based feature generator, starting with the preliminaries of the module training strategy, followed by the details of the proposed MRC architecture. Finally, we introduce the process for reconstructing the target audio.

\subsection{Text and Audio Embeddings}
\label{sec: condition}
\subsubsection{Text Embedding} For both the target text prompts and the captions of the reference audio clips, we use the pre-trained Flan-T5~\cite{t5} as the text encoder to extract the text feature. Compared to contrastive language pretraining models, such as CLIP~\cite{clip} and CLAP~\cite{clap}, the Flan-T5 encoder captures both semantic meaning~\cite{tango} and temporal structures~\cite{tclap} from textual prompts, showing excellent performance in extracting semantic information for text-to-audio generation tasks~\cite{tango,reaudioldm}. Denoting the text prompt as $\boldsymbol{c}$, the text embedding $\boldsymbol{C}$ is obtained as: 
\begin{equation}
\label{eqa:1}
    \boldsymbol{C} = \textit{f}_{\text{T5}}(\boldsymbol{c})
\end{equation}
where $\textit{f}_{\text{T5}}(\cdot)$ is the Flan-T5 text encoder~\cite{t5}. 

For each caption of the reference audio concepts, $\boldsymbol{t}_k, k=1,...,K$, %two different formats of the text feature are applied during the feature extraction and feature interaction stages, respectively. 
we first use the same Flan-T5 model to compute the embedding as $\boldsymbol{e}_k = \textit{f}_{\text{T5}}(\boldsymbol{t}_k)$. We then concatenate the embeddings of the reference captions as the reference embedding input $\boldsymbol{E}$, as follows: 
\begin{equation}
\label{eqa:1}
    % \boldsymbol{E}^{r} =[<\textit{f}_{\text{t5}}(\boldsymbol{t}_i)>,<\textit{f}_{\text{t5}}(\boldsymbol{t}_2)>,...,<\textit{f}_{\text{t5}}(\boldsymbol{t}_k)>] 
    \boldsymbol{E} =[\boldsymbol{e}_1,\boldsymbol{e}_2,...,\boldsymbol{e}_K]. 
\end{equation}
This enables the model to interact with all the text embeddings of the reference concepts.

\subsubsection{Audio Embedding} 
\begin{figure*}[htbp]
  \centering
  \includegraphics[width=0.99\textwidth]{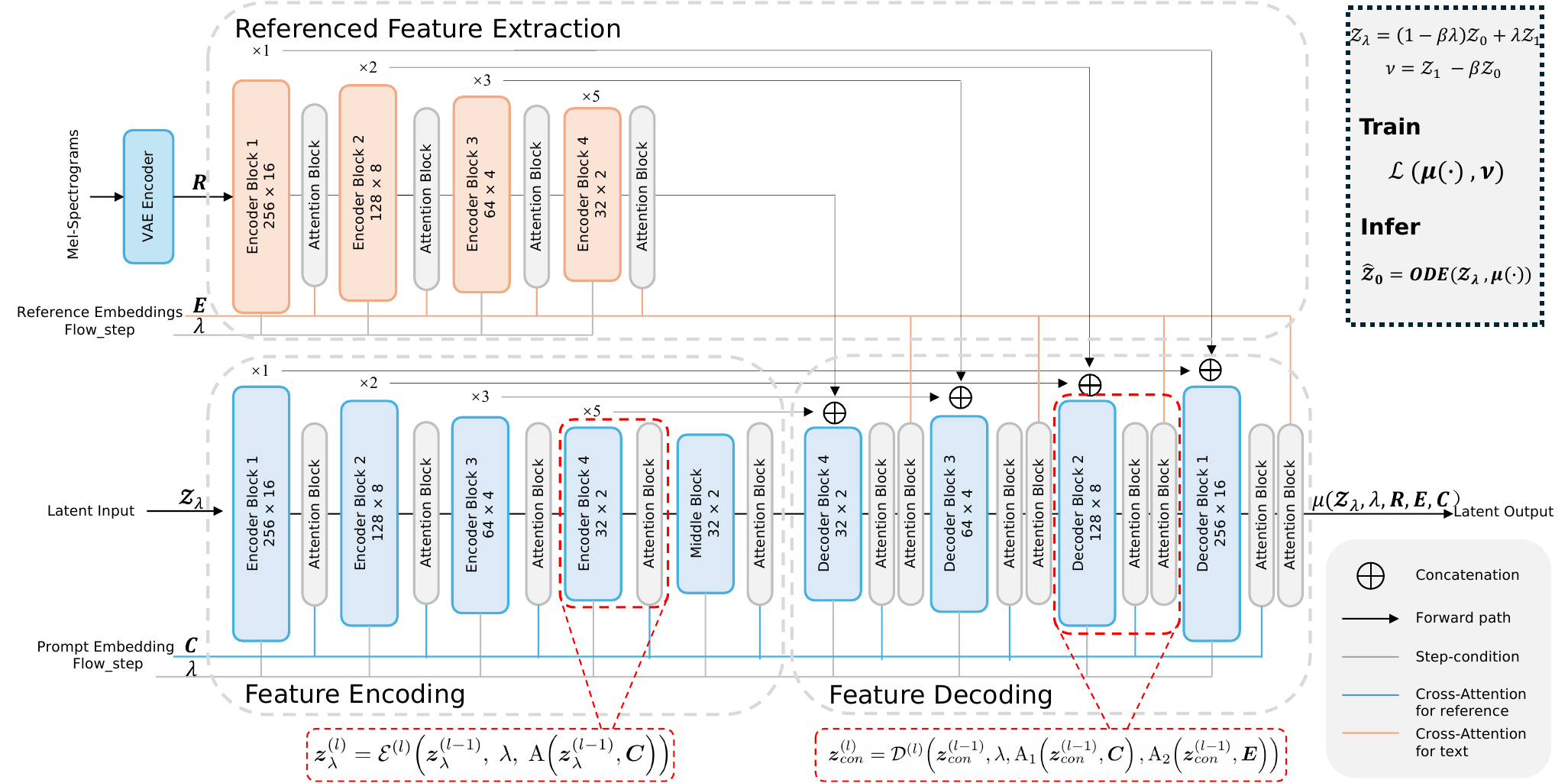}
  \caption{The details of the MRC module, which takes the reference feature $\boldsymbol{R}$ and $\boldsymbol{E}$, prompt feature $\boldsymbol{C}$ and the current noisy data $\boldsymbol{z}_{\lambda}$ as inputs to generate the dynamics for denoised data $\boldsymbol{z}_{1}$ on step $\lambda$. The output $\mu(\cdot)$ can then be used for both training and inference. }
  \label{fig:MERI}
\end{figure*}

We first follow the baseline models~\cite{audioldm} by applying a pre-trained VAE encoder $\textit{f}_{\text{VAE}}(\cdot)$ to encode the reference audio clip from the mel-spectrogram into intermediate representations within the latent space. Taking $\boldsymbol{a}_{\lambda}$ as the mel-spectrogram of the original waveform $\mathbf{a}$, the latent feature for audio is obtained as $\boldsymbol{z}_{\lambda} = \textit{f}_{\text{VAE}}(\boldsymbol{a}_{\lambda})$. %The content representation is the target feature for training the diffusion framework. 
Similarly, the latent representation for the mel-spectrogram of referenced audio $\boldsymbol{a}_k$ is obtained from the VAE encoder~\cite{audioldm}, as follows: 
\begin{equation}
    \boldsymbol{r}_k = \textit{f}_{\text{VAE}}(\boldsymbol{a}_k)
\end{equation}
%where each reference $\boldsymbol{r}_k$ is processed separately during the feature extraction stage. Then during the feature decoding stages, 
The embeddings $\boldsymbol{r}_k, k=1,...,K,$ are then concatenated into $\boldsymbol{R}$, shown as input in Figure~\ref{fig:MERI}: 
\begin{equation}
    % \boldsymbol{r} = [<\textit{f}_{\text{vae}}(\boldsymbol{a}_1)>,<\textit{f}_{\text{vae}}(\boldsymbol{a}_2)>,...,<\textit{f}_{\text{vae}}(\boldsymbol{a}_k)>]
    \boldsymbol{R} = [\boldsymbol{r}_1,\boldsymbol{r}_2,...,\boldsymbol{r}_K]
\end{equation}
% In addition, the pre-trained VAE decoder is applied to decode the updated latent vector $\boldsymbol{r}_k$ into the mel-spectrogram $\boldsymbol{a}_k$, which is applied in the final stage in Section~\ref{sec: reconstruction} to reconstruct the output waveforms.

\subsection{Audio Feature Generation}
\label{sec: diffusion}

\subsubsection{Rectified Flow Matching}

Traditional diffusion-based generative models, such as DDPMs~\cite{DDPM}, generate samples through iterative denoising from a Gaussian prior, requiring hundreds of steps to gradually turn noise into the target data. However, such approaches are highly dependent on noise schedules and can be slow in sampling. RFM~\cite{liuflow}, on the other hand, addresses the sampling issue and increases stability by learning a continuous flow between noise and data distributions, as demonstrated in audio related tasks~\cite{flowsep,audiobox}.

Unlike DDPMs that operate on discrete time steps $t \in \{0, 1, ..., N\}$, RFM models introduce a continuous flow variable $\lambda \in [0, 1]$ that smoothly interpolates between Gaussian noise $\boldsymbol{z}_0 \sim \mathcal{N}(\mathbf{0}, \mathbf{I})$ and the target data representation $\boldsymbol{z}_1$. By replacing discrete time steps with a continuous flow parameter $\lambda$, RFM enables smooth noise-to-data interpolation and a stable, schedule-free training objective based on a constant velocity field $\boldsymbol{v}$. The noisy data $\boldsymbol{z}_\lambda$ at the flow parameter of $\lambda$ is defined by a linear equation:
\begin{equation}
\boldsymbol{z}_{\lambda} = (1 - \beta\lambda)\boldsymbol{z}_{0} + \lambda\boldsymbol{z}_{1}
\end{equation}
\noindent where $\beta = (1-\sigma)$ is a small positive constant (e.g., $\sigma = 1 \times 10^{-5}$) introduced to avoid degeneracy and to improve numerical stability. This formulation allows the model to define a continuous transformation path between the noise and the target.

The model aims to estimate this velocity field $\boldsymbol{v}$, which defines the direction and magnitude of the continuous transformation that transports samples from the noise distribution to the data distribution. To train the model, a neural network $\mu(\cdot)$, corresponding to the proposed MRC module, is optimized to predict the velocity field $\boldsymbol{v} = \boldsymbol{z}_1 - \beta\boldsymbol{z}_0$.
% To train the model, a neural network $\mu(\cdot)$, which will be introduced as MRC in the next section, is optimized to predict the velocity field $\boldsymbol{v} = \boldsymbol{z}_1 - \beta\boldsymbol{z}_0$. 
Unlike denoising targets in general diffusion models, this velocity field is independent of $\lambda$, which provides a consistent supervision signal in all flow positions sampled and improves training stability. The network is trained to minimize the following objective:
\begin{equation}
L_{\text{RFM}}(\theta) = \mathbb{E}_{\lambda, \boldsymbol{z}_1,\boldsymbol{z}_0} \left\| \mu(\boldsymbol{z}_{\lambda},\boldsymbol{R},\lambda,\boldsymbol{E},\boldsymbol{C}) - \boldsymbol{v}\right\|^2
\end{equation}
Here, the network is conditioned on multiple modalities, including the reference audio features $\boldsymbol{R}$, the text embedding $\boldsymbol{E}$, prompt $\boldsymbol{C}$ and the continuous flow parameter $\lambda$, thus enabling fine-grained control over the generation.

% During inference, an ODE solver integrates the dynamics defined by the trained network $\mu(\cdot)$, starting from Gaussian noise $\boldsymbol{z}_0$ and producing $\boldsymbol{z}_1$. 

During inference, an ordinary differential equation (ODE) of the form $\frac{d\boldsymbol{z}_\lambda}{d\lambda} = \mu(\boldsymbol{z}_\lambda, \boldsymbol{R}, \lambda, \boldsymbol{E}, \boldsymbol{C})$
is solved using the numerical ODE solver~\cite{liuflow}, starting from Gaussian noise $\boldsymbol{z}_0$ and integrating from $\lambda=0$ to $\lambda=1$ to obtain the final sample $\boldsymbol{z}_1$. 
Compared to the hundreds or even thousands of iterations required in diffusion-based sampling~\cite{ddim}, this process typically requires fewer than 50 iterations to synthesize high-quality audio.

\subsubsection{Multi-Reference Customization}
\label{sec: interaction}

To produce customized audio features that incorporate referenced concepts without requiring fine-tuning, we designed the MRC structure. As shown in Figure~\ref{fig:MERI}, the MRC module merges textual and acoustic inputs, drawn from the input prompt and referenced concepts, and predicts the necessary dynamics to yield the denoised output $\boldsymbol{z}_1$.

The design draws inspiration from ControlNet~\cite{controlnet}, which adds a conditional encoder to steer the decoder. However, as illustrated in Figure 3, MRC adopts a U-Net backbone with dedicated encoder blocks in the down-sampling stage to process multiple referenced concepts. These new encoder blocks are externally linked to corresponding decoder blocks during up-sampling, enabling feature integration at multiple scales. Unlike ControlNet, which typically processes spatially aligned visual conditions (e.g., edges or depth maps), MRC handles multiple audio-text references with an external encoder. These references are fed into the generation pipeline via two separate mechanisms: extracted audio latent features are fused directly with the input of each decoder block for feature-level guidance, while text features are directed to cross-attention layers to ensure semantic alignment.

Rather than merely copying information from the references, MRC learns to identify relevant audio-event characteristics from multiple sources and blend them into the target generation. Specifically, the down-sampling stage comprises two parallel encoder paths: one for the generated feature vector and the other for the referenced feature vector. Their outputs are concatenated before being passed to a unified up-sampling decoder, ultimately producing the final latent representation.

% \noindent\textbf{In feature encoding path:} the encoder layers of the feature encoding take the noisy latent feature $\boldsymbol{z}_{\lambda}$ and the text embedding $\boldsymbol{C}$. This module operates for general latent feature encoding. 

\noindent\textbf{Feature Encoding Path.} The encoder layers take the noisy latent representation $\boldsymbol{z}_{\lambda}$ and the text prompt embedding $\boldsymbol{C}$ as input. Formally, for each encoding block $\mathcal{E}^{(l)}(\cdot)$, the latent feature is updated along with the flow parameter $\lambda$ as
\begin{equation}
\label{eqa:attn}
\boldsymbol{z}_{\lambda}^{(l)} = \mathcal{E}^{(l)}\!\left(
\boldsymbol{z}_{\lambda}^{(l-1)},\;
\lambda,\;
\mathrm{A}\!\left(\boldsymbol{z}_{\lambda}^{(l-1)}, \boldsymbol{C}\right)
\right)
\end{equation}
where $\boldsymbol{z}_{\lambda}^{(0)} = \boldsymbol{z}_{\lambda}$ denotes the input noisy latent feature, $\mathcal{E}^{(l)}(\cdot)$ represents the $l$-th encoding block, and $\mathrm{A}(\cdot)$ denotes the attention operation that injects text conditioning information $\boldsymbol{C}$ into the latent representation.

\noindent\textbf{Reference Feature Extraction Path.}
The encoder layers for reference feature extraction, on the other hand, take the latent representation of the reference audio concept $\boldsymbol{R}$ and the corresponding textual embedding $\boldsymbol{E}$. Applying the same equation (\ref{eqa:attn}), this approach allows \textit{DreamAudio} to identify and extract the audio feature for the customized content based on the reference audio and their corresponding captions. It is noted that the two groups of encoder blocks share the same structure, e.g., convolutional layers with a cross-attention module for text embedding, but are applied with independent weights for different downsampling purposes, respectively. 

\noindent\textbf{Feature Decoding Path.} We feed all the features and conditions through a group of decoder blocks, designed as the feature decoding path. Specifically, during the up-sampling stages, the latent vector from the feature encoding path and the feature extraction path are concatenated with the audio feature from each scaling level in downsampling stages via skipped connections, before passed into each decoder block. Different from single text embedding conditions in the feature encoding path, both the target prompt $\boldsymbol{C}$ and reference captions $\boldsymbol{E}$ are given as conditions through two cross-attention blocks respectively. The updated latent state for the $l$-th decoding block $\mathcal{D}^{(l)}$ is formed as:
% \begin{equation}
% \label{eqa:attn2}
% \boldsymbol{z}_{con}^{(l)} = \mathcal{D}^{(l)}\!\left(
% \boldsymbol{z}_{con}^{(l-1)}, \;  
% \lambda,\;
% \mathrm{A}_{1}\!\left(\boldsymbol{z}_{con}^{(l-1)}, \boldsymbol{C}\right)
% ,
% \mathrm{A}_{2}\!\left(\boldsymbol{z}_{con}^{(l-1)}, \boldsymbol{E}\right)
% \right)
% \end{equation}
\begin{equation}
\label{eqa:attn2}
\resizebox{0.91\linewidth}{!}{$
\boldsymbol{z}_{con}^{(l)} = \mathcal{D}^{(l)}\!\left(
\boldsymbol{z}_{con}^{(l-1)}, \lambda,
\mathrm{A}_{1}\!\left(\boldsymbol{z}_{con}^{(l-1)}, \boldsymbol{C}\right),
\mathrm{A}_{2}\!\left(\boldsymbol{z}_{con}^{(l-1)}, \boldsymbol{E}\right)
\right)
$}
\end{equation}
where $ \boldsymbol{z}_{con}^{(l-1)} = [\boldsymbol{z}_{\lambda}^{(l-1)};\boldsymbol{R}^{(l-1)}]$ is the concatenation of the latent state from both the feature encoding path and the reference feature extraction path.  $\mathrm{A}_{1}(\cdot)$ and $\mathrm{A}_{2}(\cdot)$ are the two blocks that calculate the cross-attention for the prompt embedding and the reference embedding, respectively.

\subsection{Audio Feature Reconstruction}
\label{sec: reconstruction}
As shown in Figure~\ref{fig:overview}, \textit{DreamAudio} leverages both the VAE decoder and a generative adversarial network (GAN)-based vocoder for reconstructing the audio feature from latent space into the target waveform. Followed by the RFM model, the generated audio representation is first decoded into the mel-spectrogram by the VAE decoder. In the next stage, a vocoder is applied to convert the audio feature into the waveform as the final output, and we train the vocoder using the SOTA structure BigVGAN~\cite{bigvgan} in the proposed system.

\subsection{Customized Data Processing}

In the early training stages, we observed that providing excessively detailed audio features caused the model to become overly reliant on these given representations, thereby limiting its ability to independently learn and generate new audio characteristics. To mitigate this, augmentation strategies are implemented to enhance the generalisation ability beyond the provided references. First, we randomly mask the contents of the customised audio event for the target audio. Second, we randomly remove the customised content during training. We train the DreamAudio model with a $10\%$ content masking and a $40\%$ content dropping rate. \textcolor{black}{These augmentation strategies, i.e. masking and dropping, are applied to both reference audio and reference text to improve the model's robustness.} Experiments on the effectiveness of these masking and dropping methods are given in Section~\ref{masking_dropping_exp}.

\section{Datasets and Evaluation Benchmark}

% \subsection{Dataset} 
% \noindent
\label{sec: dataset}

Previous text-to-audio generation models~\cite{audioldm,tango,sound_vecaps} mainly work on public audio-language datasets~\cite{audioset,wavcaps,autoacd} These datasets only provide audio clips and their corresponding captions. In customized generation tasks, additional reference concepts, including reference audio clips and corresponding captions are also required. To this end, we construct the customized training and evaluation datasets based on four commonly used audio-language datasets. The construction of these datasets enables \textit{DreamAudio} to extract the customized feature from specific user contents. In addition, we manually collect a small-scale dataset with real-world customized scenarios as the first benchmark for CTTA. In the following sections, we first introduce the datasets collected as the base dataset to form the customized datasets. Then, we discuss the strategies for developing the training and testing data, followed by the introduction of the benchmark dataset. All evaluation benchmarks used for the system are presented in the end. 

\begin{figure*}[htbp]
  \centering
  \includegraphics[width=1.0\textwidth]{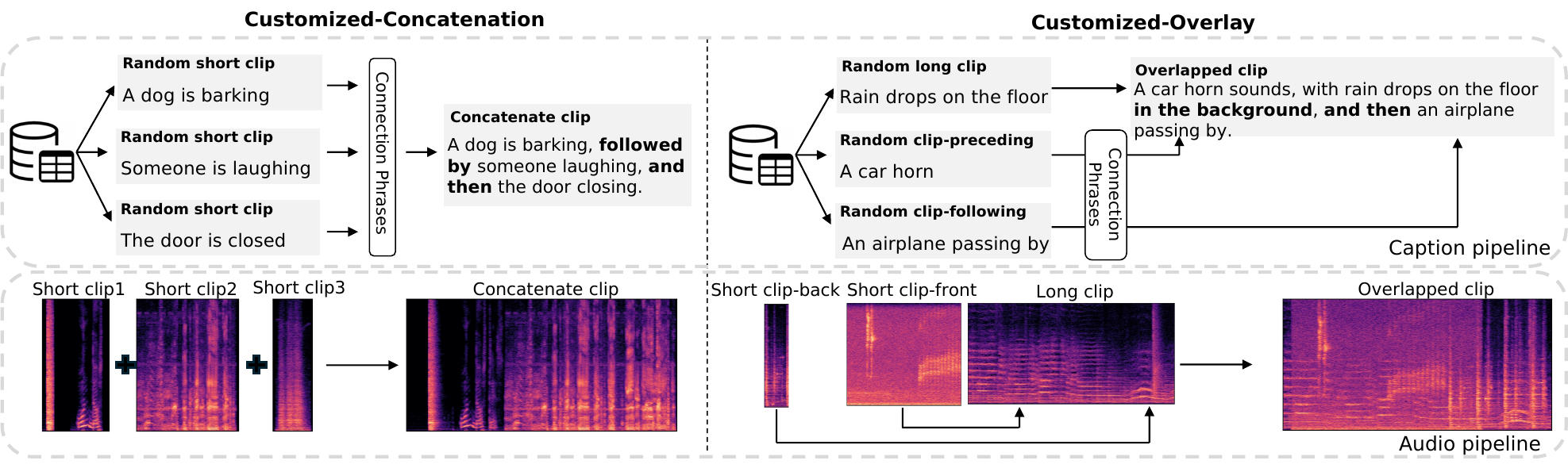}
  \caption{The generation pipeline of the customized datasets, with the Customized-Concatenation on the left and Customized-Overlay on the right. All the clips are selected randomly from the base dataset and both the concatenated clips and overlapped clips are fixed into 10-seconds.  }
  \label{fig:dataset}
\end{figure*}

\begin{table*}[t]
\caption{\textcolor{black}{The setup of four datasets used for training and testing of the proposed DreamAudio. The reference-type indicates the strategy for generating the reference concept and the reference-num represents the number of audio events for customization through the reference concept.}}
\label{tab:dataset}
\resizebox{0.98\textwidth}{!}{%
    \begin{tabular}{ccccccc}
    \toprule
    Dataset    & General-Dataset  & Customized-Content  & Train-Num& Test-Num & Reference-Type & Referenced-Num \\
    \midrule
    AudioCaps-General  & AudioCaps  & \ding{55}     & $49,502 $ & $ 928$   & Retrieval-based & \ding{55} \\
    Customized-Overlay    & WC+UB8K+ESC & \ding{51} & $146,481 $ & $ 200$ & Overlap-audio & $1-3$\\
    Customized-Concatenation     & AC+WC+US8K+ESC & \ding{51}   & $92,299 $ & $ 200$  & Concate-audio & $1-5$\\
    Customized-Fantasy     & Online-Collect & \ding{51}   & $- $ & $ 25$  & Concate-audio & $3$\\
    \bottomrule
    \end{tabular}
}
\end{table*}

\subsection{General Datasets}

\subsubsection{AudioCaps} AudioCaps~(AC)~\cite{audiocaps} is one of the largest audio datasets with hand-crafted captions. As a subset of the AudioSet~\cite{audioset}, AudioCaps contains $52,905$ $10$-second audio clips, and each clip is matched with a human-annotated caption. Taking from original videos on YouTube, AudioCaps contains various classes of audio, including music, Foley sounds~\cite{foley}, and human speech. Each audio clip has a single caption in the training set and five captions in the test set, where we only apply the first caption of each waveform to the testing data. We follow the official dataset split, and build the training set of $49,502$ audio clips and the testing set of $928$ clips. 

\subsubsection{WavCaps} WavCaps~(WC)~\cite{wavcaps} is a machine-labeled dataset with audio captions generated through Large Language Models (LLMs). WavCaps contains $403,050$ audio clips collected from various datasets such as AudioSet~\cite{audioset} and FreeSound~\cite{freesound}, providing various durations of audio clips ranging from $0.1$s seconds to $68$ seconds. To align the length of the input and target audio clips, we only collect the clips shorter than $10$ seconds, resulting in a group of $163,818$ audio clips. Then, we randomly selected $5\%$ for testing, forming a set of similar size to the AudioCaps test set. 

\subsubsection{UrbanSound8K} UrbanSound8K~(US8K)~\cite{urbansound8k} contains $8,732$ labeled audio clips and each clip is shorter than $4$ seconds. The dataset is divided into $10$ different classes, including urban noise, background sound sources, and natural sound sources. We randomly selected $732$ clips for model testing, similar to the size of the AudioCaps test set, and the remaining 8000 clips for model training.

\subsubsection{ESC-50} ESC-50~(ESC)~\cite{esc-50} has $2,000$ $5$-second audio recordings of natural sounds and domestic sounds. These samples are evenly categorized into $50$ distinct classes. To match a similar data scale to AudioCaps, we randomly chose $400$ audio clips for model testing and $1,600$ clips for model training.

\subsection{Customized Datasets}
\label{sec: customized dataset}
% Based on the four general datasets mentioned in the previous section, we developed three different pipelines to generate customized datasets for training and evaluation. An overview of each dataset is presented in Table~\ref{tab:dataset} and the pipelines for generating such datasets are illustrated in Figure~\ref{fig:dataset}. 

With the four general audio datasets mentioned above, we design three task-specific data generation pipelines to construct customized datasets for training and evaluation. Unlike standard audio augmentation techniques, these pipelines are specifically designed for the CTTA task to simulate event-level customization. An overview of each customized dataset is presented in Table~\ref{tab:dataset}, and the corresponding data generation pipelines are illustrated in Figure~\ref{fig:dataset}.

% It is noted that based on the dataset processing pipeline, the customized audio events for testing are selected as unseen audio features and the evaluation set for both Customized-Overlay and Customized-Concatenation are designed as zero-shot cases.  

\subsubsection{Customized-Concatenation} We develop the dataset by concatenating audio events. In detail, we collect audio clips shorter than $5$ seconds from the database and then randomly form a group of audio clips whose lengths sum up to $10$ seconds. The target audio is then generated by concatenating these short audio clips. Next, we construct the referenced audio by grouping the short audio clips into three sets, each set is concatenated into a single clip as the reference audio.
% An extending threshold $\alpha_m$ is proposed to add more random clips into the referenced audio samples to train the capability of AudioCustom in identifying and extracting the customized target features. 
For the reference captions, we first convert the labels from US8K and ESC-50 into short captions by simply adding verbs and subjects. Then, we generate sentences by connecting the caption of each short audio clip with a connection phrase randomly selected from our connection list. For the Customized-Concatenation dataset, we collect $92,300$ audio-text pairs for model training and $200$ pairs for model testing.

\subsubsection{Customized-Overlay} We found that the simple concatenating strategy led to several limitations. First, not all audio events can be connected smoothly, which may leave a large blank space in the synthesized waveforms. Second, audio created solely by concatenation often sounds unnatural, failing to accurately reproduce user-provided reference contents, and thereby not reflecting realistic scenarios. To guide the model with more realistic cases, we develop a second version of the customized dataset, called Customized-Overlay dataset. Specifically, for each audio sample, we take a $10$-second audio clip as the base waveform and randomly select two shorter clips~(less than $5$ seconds) as the ``preceding'' and ``following'' audio clips. The target audio clip is then generated by adding the ``preceding'' and ``following'' audio events into the base waveform. Then, we remix the waveform of two extended clips and the base clips under a random signal to noise ratio (SNR) between -15 and 15 dB. The captioning strategy is similar to the approach we apply in the Customized-Concatenation dataset. We generated $146,481$ pairs for training and $200$ pairs for testing in the Customized-Overlay dataset.

\subsubsection{Customized-Fantasy} We manually created a small-scale dataset as a benchmark for this new task, named as Customized-Fantasy. All the data clips are collected from online sources, including many unique and special events extracted from games or movies to simulate the user cases for real-world audio-production scenarios. For example, customization in \textit{laser guns}, \textit{monster roars} and \textit{Minion speaks}. In addition, human interactions are involved to ensure that the captions are meaningful and present semantic coherence. In total, we collected 60 different events from Pixabay\footnote{\url{https://pixabay.com/}}, a website which provides royalty-free audio resources under the Pixabay License suitable for research purposes. Together, we manually developed $25$ different scenarios to simulate real-world customized generation.

\subsubsection{AudioCaps-General} To maintain the capability of the system in more robust situations, e.g., prompting on audio events without the reference concepts, we apply the general AudioCaps dataset with empty reference concepts. For instance, the model is given with the target audio clip and the target caption, along with three empty waveform and empty text as the reference concepts. The same training-testing split is applied for training and evaluation of the model. 

% In detail, we follow the pipeline from Re-AudioLDM~\cite{reaudioldm} and construct the dataset from AudioCaps. To develop the AudioCaps-Retrieval, we apply a CLAP-score function to find the Top-3 relevant audio clips by calculating the similarity between the text embedding of the target prompt and the audio embedding of the audio candidates. For each target audio clip, we provide three retrieval audio samples with their corresponding captions as the reference information. AudioCaps-Retrieval follows the official train-test split and it is noted that testing data can only be retrieved for evaluation, avoiding early access to the testing data during the training stage.

\subsection{Evaluation Metrics}
Following the evaluation protocol of baseline audio generation models~\cite{audioldm2}, we use three different metrics for performance evaluation, including the Fr\'echet Audio Distance~(FAD), Kullback-Leibler~(KL) divergence and Contrastive Language Audio Pretraining~(CLAP) related score. In addition, we follow the evaluation pipelines from the customized image generations~\cite{dreambooth,freecustom} and incorporate subjective assessments that measure the consistency and the fidelity between generated and target audio, including CLAP audio score~(CLAP$_A$), AudioBox Aesthetics, and subjective evaluations. Specifically, audio-text latent similarity based metrics including FAD, KL and CLAP-score are applied for evaluating the performance on Customized-Concatenation, Customized-Overlay and general AudioCaps testing sets, subjective evaluations and AudioBox Aesthetics are used for the evaluations on Customized-Fantasy.

\subsubsection{FAD} FAD first computes the multivariate Gaussian of two groups of embedding values from a pre-trained VGGish~\cite{vggish} feature extraction model. Then, it computes the Frechet distance between the Gaussian mean and variance of two sets of high-dimensional features. A lower FAD score indicates that the generated audio group presents a closer data distribution to the target audio group based on audio features. 

\subsubsection{KL Divergence} KL measures the logarithmic difference between the probabilities assigned by the distributions of two audio clips across all possible events. We apply the audio classification model PANNs~\cite{kong2020panns} for feature extraction and calculate the KL score between each target audio sample and their corresponding outputs. A lower KL represents a smaller distance in distributions, suggesting that the generated outputs are more similar to the target clips. 

\begin{table}[t]
\textcolor{black}{
    \caption{The setup of the primary experiments we performed, with DDPM for using DDPM-based training strategies, UNet for basic diffusion-based structure, and DiT for diffusion-transformer based backbone. \textit{DreamAudio} with dual-encoder and single-encoder indicates whether the two groups of encoder blocks share the same weights.}
    \label{tab: model-setups}
    \resizebox{0.5\textwidth}{!}{%
    \begin{tabular}{cccc}
    \toprule
    Model            & Param(train) &    Diffusion    &  MRC   \\
    \midrule
    DreamAudio-DDPM  & $760$M  &  DDPM   & Dual-Encoder   \\
    DreamAudio-UNet  & $760$M  &  RFM   & \ding{55}   \\
    DreamAudio-DiT-B  & $920$M  &  RFM   & Dual-Encoder   \\
    DreamAudio-DiT-L  & $1.3$B  &  RFM   & Dual-Encoder  \\
    DreamAudio-SEncoder  & $815$M  &  RFM   & Single-Encoder   \\
    DreamAudio  & $891$M  &  RFM   & Dual-Encoder   \\
    DreamAudio-L & $1.1$B  &  RFM   & Dual-Encoder   \\
    \bottomrule
    \end{tabular}
}
}
\end{table}

\subsubsection{CLAP} The CLAP score calculates the cosine similarity between the text embedding and the audio embedding provided by the CLAP model~\cite{clap}. The CLAP model learns projectors to align the audio and text embeddings into a joint space and presents paired audio and language data with similar features within the latent space. Given both generated audio embedding  $\mathbf{e}_{a}$ and text embedding $\mathbf{e}_{t}$, the score is calculated as:
\begin{equation}
    {\text{CLAP}(\mathbf{e}_{a}, \mathbf{e}_{t}) = \frac{\mathbf{e}_{a} \cdot\mathbf{e}_{t}}{\text{max}(\| \mathbf{e}_{a} \| \| \mathbf{e}_{t} \|, \epsilon)},}
\end{equation}
where $\epsilon$ is a small value to avoid zero division. The CLAP score illustrates the correspondence between the generated audio and text prompt, and a higher score shows better performance on the semantic level. 

\subsubsection{CLAP$_A$} Instead of calculating the similarity between audio and text embeddings, the CLAP$_A$ score is specially designed for the CTTA task to compare the customized concepts between the targets and generated audio by calculating the cosine similarity score between the target audio embedding $\mathbf{e}_{a}$ and the generated audio embedding $\mathbf{e}_{\hat{a}}$,
\begin{equation}
    {\text{CLAP}_{A}(\mathbf{e}_{a}, \mathbf{e}_{\hat{a}}) = \frac{\mathbf{e}_{a} \cdot\mathbf{e}_{\hat{a}}}{\text{max}(\| \mathbf{e}_{a} \| \| \mathbf{e}_{\hat{a}} \|, \epsilon)}.}
\end{equation}
\noindent A higher score means that the generated output is more similar to the target audio within the embedding space. 

\subsubsection{AudioBox Aesthetics} AudioBox Aesthetics is a recently proposed audio quality assessment tool that has demonstrated a strong correlation with subjective evaluations by human listeners~\cite{Audiobox_Aesthetics}. In detail, this metric applies a transformer-based model for extracting and assessing the quality of audio clips without reference clips. In this paper, we choose the production quality~(PQ) and content usefulness~(CU) for the task. In detail, PQ is an objective aspect of the overall quality and CU is a subjective axis to evaluate the likelihood of audio sample for content creation. 

\subsubsection{Subjective Evaluation} We adopt the evaluation metrics used in the baseline models~\cite{audioldm2}, incorporating both Overall Impression (OVL) and Audio-Text Relation (REL) for subjective assessments. Detailed scoring instructions, including illustrative examples, are provided to the raters to ensure clarity. For OVL, we follow the approach outlined in the baseline models~\cite{audioldm2} by asking raters: \textit{How would you rate the overall quality of this audio sample?}. Responses are scored on a Likert scale, ranging from 5 for ``excellent'' to 1 for ``bad''. For REL, we adapt the metric to better suit the customized generation tasks by asking: \textit{Does the generated audio successfully present the target audio events customized by the reference audio samples?} The subjective evaluation was conducted with ten human raters recruited through an open advertisement within the Department of Electrical and Electronic Engineering at the University of Surrey. All participants were PhD students and were not involved in the development of the proposed model or the writing of this paper, and none of the authors participated in the evaluation. To encourage diverse perspectives and reduce potential bias, the participant pool included six raters with backgrounds in audio related research and four raters from unrelated fields such as computer vision and robotics, providing complementary non-expert assessments.

% We use Amazon Mechanical Turk\footnote{\url{https://requester.mturk.com/}}, a crowd-sourced platform, to evaluate subjective metrics including OVL, REL, and MOS. The instructions on how to perform evaluation are clearly illustrated for the raters with examples. \textcolor{black}{Specifically, for OVL, raters were asked \textit{How would you rate the overall quality of this music? Consider its resemblance to real-world audio and its naturalness,} with a five-point scale ranging from \textit{5-Excellent quality} to \textit{1-Bad quality}. Similarly, for REL, the question posed was, \textit{How would you rate the relevance of music to the text description?} with a similar five-point scale for responses. In evaluating MOS, the question was, \textit{How natural does this recording sound? Take into account emotion, prosody, and other human-like details,} with options ranging from \textit{completely unnatural speech} to \textit{perfectly natural speech}.} To ensure the credibility of the evaluation result, we set requirements for the crowd-source worker with a minimum average approval rate of $60\%$ and with at least $50$ approvals in the record. 
% Each audio clip is evaluated by at least $10$ different raters. All three subjective metrics have a Likert scale~\cite{likert1932technique} between one and five, where a larger number indicates better performance. Study raters received payment at or above the US minimum wage. We average the scores among all raters and samples as the final score for a system. 

\begin{table*}[tbp]
\centering
\small
\caption{Comparison of model performances on the customized-based evaluation set. AC is short for AudioCap dataset, CM is for Customized-Concatenation and CE is for Customized-Overlay.}
\resizebox{0.98\textwidth}{!}{

\begin{tabular}{cccccc|cccc|cccc}
\toprule
\multirow{3}{*}{Model}          & \multirow{3}{*}{Dataset} &  \multicolumn{4}{c}{Customized-Concatenation} & \multicolumn{4}{c}{Customized-Overlay} & \multicolumn{4}{c}{Customized-Fantasy} \\
 \cmidrule(lr){3-14}
&  &FAD $\downarrow$ & KL $\downarrow$   &CLAP $\uparrow$ & CLAP$_{A}$ $\uparrow$ &FAD $\downarrow$ & KL $\downarrow$   &CLAP $\uparrow$ & CLAP$_{A}$ $\uparrow$ &PQ $\uparrow$ & CU $\uparrow$ &OVL $\uparrow$ & REL $\uparrow$\\
\midrule

Re-AudioLDM           & AudioCaps            &   $3.05$              &        $3.24$       &         $39.4$       & $48.7$ & $2.96$ &$3.09$ & $44.3$ & $47.8$ & $5.80$ & $5.05$ & $3.15$ & $3.26$\\
\midrule

DreamAudio-DDPM    & AC+CM+CE        & $0.94$                 & $1.09$                 & $52.3$                &  $79.8$                       &           $0.99$ & $1.32$  &   $\mathbf{46.7}$& $83.3$ & $6.18$ & $5.45$ & $3.70$ & $3.53$\\
DreamAudio-UNet   &  AC+CM+CE             &   $1.59$              &        $1.73$       &         $48.1$       & $73.5$ & $1.03$ &$0.87$ & $41.8$ & $77.8$ & $5.59$ & $4.78$ & $3.41$ & $3.39$ \\
DreamAudio-DiT-B   &  AC+CM+CE             &   $0.84$              &        $1.18$       &         $50.5$       & $78.9$ & $0.94$ &$1.05$ & $41.9$ & $79.3$ & $6.22$ & $5.39$ & $3.55$ & $3.51$ \\
DreamAudio-DiT-L   &  AC+CM+CE             &   $0.49$              &        $0.95$       &         $52.1$       & $84.9$ & $0.81$ &$0.90$ & $43.5$ & $82.1$ & $\mathbf{6.42}$ & $\mathbf{5.85}$ & $3.79$ & $3.85$ \\
DreamAudio-SEncoder   &  AC+CM+CE             &   $0.79$              &        $1.05$       &         $51.4$       & $85.9$ & $0.88$ &$0.92$ & $41.5$ & $81.3$ & $6.12$ & $5.31$ & $3.51$ & $3.69$  \\
\midrule
DreamAudio    &  AC+CM+CE           &   $0.50$              &        $\mathbf{0.90}$       &         $52.0$       & $86.3$ & $0.78$ &$0.82$ & $42.0$ & $81.7$ & $6.31$ & $5.56$ & $3.74$ & $3.91$ \\
DreamAudio-L    &  AC+CM+CE          &   $\mathbf{0.46}$  &  $0.92$  & $\mathbf{52.5}$ & $\mathbf{87.7}$ & $\mathbf{0.73}$ &$\mathbf{0.67}$ & $42.4$ & $\mathbf{83.9}$ & $6.37$ & $5.65$ & $\mathbf{3.89}$ & $\mathbf{4.17}$    \\

\midrule
\end{tabular}
}
\label{tab: customized-generation}
\end{table*}

\section{Experimental Setting}

\subsection{Model Architecture Details}

 For the text encoder, we use a pre-trained Flan-T5 model~\cite{t5}, resulting in a $1024$ dimension feature sequence with a fixed length of $50$ for every caption. We use the pre-trained VAE model from AudioLDM~\cite{audioldm} which provides a compression ratio of $4$ and results in a latent vector of $256$ dimension in temporal and $16$ dimension in frequency for a ten-second mel spectrogram. For the vocoder, we apply BigVGAN~\cite{bigvgan} and perform the self-supervised pre-training on audio clips sampled at $16$ kHz. Table~\ref{tab: model-setups} summarizes our experiments. We perform the experiments with two sizes of the latent diffusion model, \textit{DreamAudio} and \textit{DreamAudio-L}, with the hidden dimension size of $n_{\text{hidden}} = 96$ and $n_{\text{hidden}} = 128$, respectively. In addition, we evaluated several variants of the proposed system. The DreamAudio trained under a DDPM framework is denoted as \textit{DreamAudio-DDPM}. \textcolor{black}{The version using a standard UNet backbone but without the MRC module is named \textit{DreamAudio-UNet}. In this system, rather than using the dual-path MRC structure, all three reference audio latent features are concatenated together with the latent of the noisy input at the input layer, extending the input dimension to align with the MRC process.} The system with shared-weight MRC encoders is denoted as \textit{DreamAudio-SEncoder}, where the encoder blocks used for referenced feature extraction and target feature encoding share the same encoder. This encoder is simply run multiple times for the target latent and each referenced latent during training and inference. Beyond the UNet-based architectures, we also experimented with two models that adopt a diffusion-transformer (DiT) backbone~\cite{dit}. In \textit{DreamAudio-DiT-B}, we employ 12 transformer layers, while in \textit{DreamAudio-DiT-L}, we use 24 transformer layers. Both models incorporate skip connections, and first half of the layers are used as the encoders while the second half is applied as the decoders.

\subsection{Training and Inference Setup}
Despite the pre-trained text encoder and VAE, we train the model and vocoder separately. We first train the vocoder on a large-scale audio dataset for $1$M steps and then freeze the module while we train the proposed \textit{DreamAudio}. We collected $288,283$ clips for model training and $1,328$ clips for model evaluation by combining Customized-Concatenation, Customized-Overlay, and general AudioCaps datasets mentioned in Section~\ref{sec: customized dataset}. For each audio clip, we provide three-pairs of reference concepts as the external data. The system is trained on a single NVIDIA A100 80GB GPU for $2$ millions steps. Following similar strategies from generative networks, we apply the Classifier Free Guidance~(CFG) for RFM with a value of $2.0$ for both \textit{DreamAudio} and \textit{DreamAudio-L}. We utilize the AdamW~\cite{loshchilov2017decoupled} optimizer with a learning rate of $5\times10^{-5}$ for training and apply a linear warming up for $10,000$ steps.

\section{Results and Analysis}

We evaluated DreamAudio on both customized and general text-to-audio generation. The following section first discusses the performance of our system on the CTTA tasks and then explores the capability of general TTA tasks. 

\subsection{Customized Text-to-Audio Generation}
 All the models in Table~\ref{tab: customized-generation} are evaluated on both Customized-Concatenation and Customized-Overlay datasets. We compare our method with Re-AudioLDM~\cite{reaudioldm} as this is the only work related to customized and zero-shot tasks. The result for AudioBox-TTA-RAG~\cite{audiobox-tta} is not included because the authors did not provide any model or checkpoint. For Re-AudioLDM, we apply the system with the retrieval number of $3$ as the retrieval-based information.  As shown in Table~\ref{tab: customized-generation}, the proposed \textit{DreamAudio} significantly outperforms the previous systems across most of the metrics. The baseline system, Re-AudioLDM, achieves a FAD score of $3.05$ and $2.96$ for Customized-Concatenation and Customized-Overlay, respectively. Our systems, on the other hand, illustrate a substantially enhanced score of $0.46$ for the concatenated testing set and $0.73$ for the overlapped testing set. \textit{DreamAudio-L} also achieves the best KL divergence score of $0.92$ and $0.67$ on the two datasets, respectively. 
 
 For the CLAP score, our system demonstrates the best performance on the Customized-Concatenation dataset and achieves comparable results on the Customized-Overlay dataset. Unlike other objective metrics that directly compare target and generated audio samples, the CLAP score measures semantic alignment rather than precisely evaluating the presence of customized audio content. In this case, it is less effective in evaluating the CTTA task. In contrast, the CLAP$_A$ score evaluates the similarity between the features from each generated audio clip and its corresponding target audio clip, making it a more accurate metric for reflecting the correctness of customized content. \textit{DreamAudio-L} achieves a CLAP$_A$ score of $87.7$ on the Customized-Concatenation dataset and $83.9$ on the Customized-Overlay dataset, significantly surpassing all previous models. For the Customized-Fantasy testing set, \textit{DreamAudio} achieves the best performance among production quality, content usefulness and human evaluation, where \textit{DreamAudio-L} achieves better results on OVL of $3.89$ and REL of $4.17$. The significant enhancement of these metrics illustrates that DreamAudio can generate specific audio features related to the reference concepts and faithful to text prompts.

 \begin{table}[tbp]
\centering
\small
\caption{Comparison of model performances on the AudioCaps evaluation set. AC is short for AudioCaps dataset, CC is for Customized-Concatenation and CO is for Customized-Overlay. The dataset marked with $\ast$ indicates fine-tuning and $^\dagger$ means training from scratch on AudioCaps. }
\resizebox{0.5\textwidth}{!}{
\begin{tabular}{lcccccccc}
\toprule
Model              & Dataset & FAD $\downarrow$           & KL $\downarrow$            & CLAP $\uparrow$   & CLAP$_{A} \uparrow$   \\
\midrule
AudioLDM           & AC+AS+2 others           & $5.25$          & $1.90$          & $42.1$          & $53.5$ \\
AudioGen          & AC+AS+8 others           & $2.87$          & $1.52$          & $46.4$          & $60.2$ \\
Make-an-Audio          & AC+AS+13 others            & $2.39$          & $1.64$          & $45.4$          & $59.8$ \\
Stable Audio Open    &  FreeSound    & $4.05$ & $2.11$ &   $44.9$   &      $55.2$   \\
Tango         & AudioCaps            & $2.24$          & $1.04$          & $51.7$          & $66.2$ \\
AudioLDM2         & AC+AS+6 others            & $2.56$          & $1.75$          & $45.8$          & $55.5$ \\
TangoFlux         & WavCaps+2 others            & $2.33$          & $1.09$          & $50.4$          & $59.5$ \\
Re-AudioLDM           & AudioCaps             & $1.85$          & $1.46$          & $49.9$          & $62.0$ \\
\midrule
DreamAudio    &  AC+CC+CO    & $4.25$ & $2.48$ &   $34.9$   &      $43.6$   \\

%\textit{DreamAudio-L}  &  AC-Re+CM+CE   & $4.00$ & $2.73$ &  $35.6$ &      $44.2$  \\
DreamAudio    &  AudioCaps$^\ast$        & $1.92$ & $1.51$ & $47.5$                &  $58.8$   \\
DreamAudio    &  AudioCaps$^\dagger$     & $1.90$ & $1.50$ & $50.8$                &  $62.5$   \\
% DreamAudio-DDPM    &  AudioCaps-Re$^\ast$        & $1.61$                 & $1.52$   & $47.5$                &  $58.8$                 \\
\midrule
\end{tabular}
}
\label{tab: general-generation}
\end{table}

\subsection{General Text-to-Audio Generation}
In this section, we evaluate the capability of \textit{DreamAudio} on general text-to-audio generation tasks. We compare the performance on audio generation with several SOTA systems, including AudioLDM~\cite{audioldm}, AudioGen~\cite{audiogen}, Make-an-Audio~\cite{makeanaudio2}, Stable Audio Open~\cite{stableaudio},  Tango2~\cite{tango}, AudioLDM2~\cite{audioldm2}, TangoFlux~\cite{tangoflux}, and Re-AudioLDM~\cite{reaudioldm}. All the checkpoints are downloaded from open-sourced GitHub or HuggingFace to reproduce the results.  For plain text-to-audio tasks without any reference inputs, DreamAudio is provided with empty reference audio clips and empty reference text token.
% As shown in Table~\ref{tab: general-generation}, DreamAudio trained on these results demonstrate that our \textit{DreamAudio} can maintain good performance in general text-to-audio tasks even without access to the corresponding customized content. 
As shown in Table~\ref{tab: general-generation}, DreamAudio trained on reference-based data does not achieve SOTA performance when directly applied to general text-to-audio generation tasks. We believe the main reason for this is that DreamAudio is architecturally designed for customized generation with reference-based conditions. The MRC module provides an additional pathway for extracting and leveraging external audio/text reference signals. When reference data are available during training stages, the model learns to rely on these external concepts for capturing acoustic feature and fine-grained attributes. This makes the model less dependent on the text conditions while generating audio. Consequently, when operating on general tasks without reference concepts, the external pathway of MRC is un-used. Hence, the performance is limited since the model has been guided by the conditions from both reference concepts and text prompts, rather than purely text prompts.

After fine-tuning DreamAudio on AudioCaps, the model recovers strong performance and achieves comparable results to most non-retrieval based systems. Furthermore, when DreamAudio is trained from scratch with only AudioCaps and no reference-based data, the model exhibits even better performance, demonstrating that the architecture itself remains fully capable of high-quality general text-to-audio generation. These findings indicate that the system can be applied to general text-to-audio generation tasks by reducing the conditioning of the reference concepts.

\begin{figure}[t]
  \centering
  \includegraphics[width=0.35\textwidth]{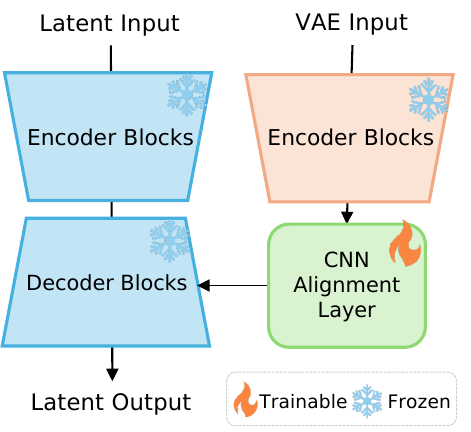}
  \caption{The details of the MRC UNet network for reference length fine-tuning. All the existing blocks are frozen and only the introduced CNN alignment layer is trained during this stage. }
  \label{fig:various_length}
\end{figure}

\begin{table}[ht]
\caption{Experimental results for customized generation with four reference concepts, where the models are obtained by fine-tuning the model trained with $3$ references. The step number with $^\ast$ indicates that the model is fully-trained with $3$ references and without any further fine-tuning.}
\label{tab: various-length}
\resizebox{0.48\textwidth}{!}{%
\begin{tabular}{cccccc}
\toprule
Reference Num              & Steps & FAD $\downarrow$           & KL $\downarrow$            & CLAP $\uparrow$   & CLAP$_{A}$ $\uparrow$   \\
\midrule
3   &  $0^\ast$    & $0.50$ & $0.92$ &   $52.5$   &      $87.7$    \\
4   &  $0^\ast$    & $14.31$ & $6.01$ &   $34.9$   &      $43.6$    \\
4    &  $10,000$  & $3.86$ & $3.01$ &  $45.6$ &      $69.5$   \\
4   &  $50,000$   & $0.79$ & $1.15$ & $49.2$ &   $77.5$ \\

\bottomrule
\end{tabular}
}
\end{table}

\subsection{Customized Generation on Various Number of References}

Our DreamAudio is primarily trained and evaluated with three-pairs of reference concepts. In this section, we investigate the capability of \textit{DreamAudio} to perform customized generation with four pairs of reference concepts. 
The current backbone adopts a CNN-based U-Net architecture, where all intermediate latent feature maps within the decoder must maintain fixed spatial dimensions to allow their concatenation through skip connections. In this case, the referenced VAE latent features need to be projected into a fixed-size representation, e.g. $96$ in the temporal dimension and $6$ in the frequency dimension for three reference pairs. When the system is given more than three-pairs of reference concepts, the dimension of reference latent feature provided by the feature extraction path does not match with the designed input. To address this issue, we introduce a lightweight CNN alignment layer that maps the reference VAE inputs with variable temporal lengths into this fixed-size latent space, as shown in Figure~\ref{fig:various_length}.

During fine-tuning, all backbone encoder–decoder blocks are frozen and only the CNN alignment layer is trained to adapt to the scenarios with different numbers of reference concepts. The performance results of \textit{DreamAudio} trained or fine-tuned with four-pairs referenced concepts are shown in Table~\ref{tab: various-length}. The first row shows the performance of a fully-trained DreamAudio model using 3 reference concepts. In contrast, a model using 4 references without any fine-tuning (0 steps) fails, as its untrained CNN alignment layer cannot properly reshape or compress the four latent inputs into a compatible, fixed-size representation. 
This results in meaningless outputs and a sharp drop in performance. However, after just 10,000 fine-tuning steps, the alignment layer learns to effectively project the four references into a unified latent space, adapting quickly to the additional input. Performance continues to improve with extended training, achieving a competitive FAD score of 0.79 after 50,000 steps. These results demonstrate that DreamAudio can readily generalize to different numbers of reference concepts by integrating a simple, trainable CNN alignment layer.

\begin{table*}[tbp]
\centering
\small
\caption{Ablation studies on data augmentation (masking, dropping) and the effectiveness of different proportions of training data. The general task is evaluated on the AudioCaps testing set and CTTA is evaluated on the Customized-Concatenation and Customized-Overlay testing sets. For data processing, AC is short for AudioCaps, and CC and CO are short for Customized-Concatenation and Customized-Overlay. The number indicates the quantity of each dataset for training. The result with $^\dagger$ indicates the experiments without reference text concepts.  }
\resizebox{0.98\textwidth}{!}{
\begin{tabular}{cccccc|cccc|cccc}
\toprule
\multirow{3}{*}{Model}     &    \multicolumn{5}{c}{Training Data-Mixing} & \multicolumn{4}{c}{General Text-to-Audio Task} & \multicolumn{4}{c}{Customized Text-to-Audio Task} \\
 \cmidrule(lr){2-14}
 & AC &CC  & CO  & Masking  & Dropping &FAD $\downarrow$ & KL $\downarrow$  &CLAP $\uparrow$ & CLAP$_{A}$ $\uparrow$ &FAD $\downarrow$ & KL $\downarrow$   &CLAP $\uparrow$ & CLAP$_{A}$ $\uparrow$\\
\midrule
 \multirow{8}{*}{DreamAudio}    & \multirow{4}{*}{$49$K} & \multirow{4}{*}{$92$K} & \multirow{4}{*}{$146$K} 
  & $10\%$ & $40\%$ & $4.25$ & $2.48$ & $34.9$ & $43.6$ & $0.64$ & $0.86$ & $47.0$ & $84.0$  \\
&  & & & $0\%$ & $0\%$ & $5.28$ & $2.65$ & $31.2$ & $40.4$ & $\mathbf{0.51}$ & $\mathbf{0.56}$ & $47.7$ & $\mathbf{88.1}$ \\
&  & & & $10\%$ & $10\%$ & $4.28$ & $2.25$ & $35.2$ & $44.4$ & $0.56$ & $0.71$ & $47.2$ & $84.3$ \\
&  & & & $50\%$ & $90\%$ & $3.81$ & $2.15$ & $38.1$ & $46.9$ & $1.32$ & $1.68$ & $42.5$ & $74.3$ \\
&  & & & $-$ & $100\%^\dagger $ & $3.69$ & $2.18$ & $40.3$ & $47.5$ & $3.11$ & $2.15$ & $39.6$ & $66.8$ \\
 \cmidrule(lr){2-14}
&$24$K & $92$K & $146$K & \multirow{4}{*}{$10\%$} & \multirow{4}{*}{$40\%$} & $5.58$ & $2.69$ & $30.1$ & $39.2$ & $0.64$ & $0.88$ & $47.2$ & $84.9$ \\
&$49$K & $24$K & $24$K & &  & $\textbf{2.85}$ & $\textbf{1.82}$ & $\textbf{41.5}$ & $\textbf{51.1}$ & $2.11$ & $1.96$ & $41.9$ & $68.5$ \\
&$49$K & $92$K & $24$K & & & $5.14$ & $2.68$ & $33.9$ & $42.1$ & $0.48$ & $0.61$ & $\mathbf{47.8}$ & $86.5$ \\
&$49$K & $24$K & $146$K & & & $4.03$ & $2.11$ & $36.8$ & $44.6$ & $0.88$ & $1.01$ & $44.5$ & $80.2$ \\
\bottomrule
\end{tabular}

}
\label{tab: ablation-generation}
\end{table*}

\subsection{Ablation Studies}

In order to validate our design choice of the proposed system, a series of ablation experiments were conducted on each proposed technique within \textit{DreamAudio}.

\subsubsection{Effectiveness of RFM}

As shown in Table~\ref{tab: customized-generation}, we compare the outputs of \textit{DreamAudio} with different training approaches on various datasets. Experiments on RFM-based systems present enhanced performance on customized tasks, i.e., Customized-Concatenation and Customized-Overlay. On the other hand, we observe that \textit{DreamAudio-DDPM} trained with diffusion-based techniques offers better results in the semantic feature than \textit{DreamAudio} by achieving better CLAP scores. These experimental results demonstrate two key findings: a) RFM-based models show promising performance in generating customized and personalized content; b) DDPM-based approaches guide the models toward more robust performance by preserving the capability of generating features based on semantic information.  

\subsubsection{Effectiveness of UNet Backbone}

We also compare our UNet architecture with diffusion-transformer (DiT) models, as DiT-based backbones have been shown to achieve superior performance in large-scale diffusion systems~\cite{dit}. As reported in Table~\ref{tab: customized-generation}, under a similar number of trainable parameters, \textit{DreamAudio-DiT-B} does not achieve better performance than the UNet-based DreamAudio. By increasing the model size to approximately $1.3$B, \textit{DreamAudio-DiT-L} provides better results, however, it results in significantly longer training and inference time. These observations are consistent with previous findings that DiT architectures tend to outperform UNet models only at considerably larger scales (e.g., DiT-XL)~\cite{dit}. Based on the current results, we adopt the UNet backbone in DreamAudio as a practical trade-off between performance and computational efficiency. 

\subsubsection{Effectiveness of MRC}
We conducted various experiments with and without the external feature extraction path from the MRC module in Table~\ref{tab: customized-generation}. For \textit{DreamAudio-UNet}, the input of the generator module is a concatenation of general latent input and three VAE latent features of the reference audios. The results show that DreamAudio with only basic UNet structures exhibits significant performance degradation in all metrics. In addition, the comparison between \textit{DreamAudio-SEncoder} and \textit{DreamAudio} indicates that the use of two groups of encoder blocks for encoding and feature extraction helps improve the overall performance.

\subsubsection{Effectiveness of Data Masking and Dropping}
\label{masking_dropping_exp}
We also conducted experiments on reference concept masking, dropping, and the impact of different data proportions. As shown in Table~\ref{tab: ablation-generation}, the masking and dropping ratios significantly influence the model's performance on customized tasks. When provided with more precise reference concepts, the model learns more effectively. However, based on the KL divergence and CLAP$_{A}$ scores, we observe that the generated results contain more similar information to the target audio. This may be because the model is led to pay more attention to extraction and concatenation rather than generation, which may explain the reason of performance drop in general TTA generation tasks. This indicates that without masking and dropping, the model's performance is degraded.  
Furthermore, in the data distribution comparison, we found that an excessive amount of customized-concatenation data makes the model overly reliant on the content of the reference concept. In contrast, Customized-Overlay data helps maintain a balance between leveraging the reference concept and the text prompts. 

Training with AudioCaps data improves the robustness of the model in general TTA scenarios. However, an excessive amount of AudioCaps data causes the model to shift focus away from the features provided by the reference concept, leading to degraded performance on the CTTA task. \textcolor{black}{Lastly, the experiment with 100\% dropping represents the system trained without any reference text concepts while maintaining access to the reference audio clips, evaluating the ability of the model to perform customization based solely on the reference audio clips.} The result shows a decrease in customization performance, which illustrates the importance of reference text in providing semantic grounding and helping the model exploit the reference audio features. In addition, as shown in Table~\ref{tab: general-generation}, DreamAudio offers improved performance on general text-to-audio tasks on the AudioCaps benchmark, giving lower FAD and higher CLAP scores, as compared to AudioLDM and Stable Audio Open. This suggests that without the constraints of the reference text, DreamAudio tends to reduce its reliance on reference concepts and learn to generate audio based primarily on the target prompt.

\subsection{Limitation}
Despite giving the satisfying performance, our method still has certain limitations. 
\subsubsection{Fixed Reference Format} The requirement for reference concept on both customized audio and related captions can be challenging to satisfy in real-world applications. Preparing suitable captions for referenced audio samples can lead to extra workloads and pose a practical barrier to general users.
\subsubsection{Fixed Audio Length} Our current architecture is primarily developed for 10-second audio, aligned with those in training data. Although the model can be extended to generate longer clips (e.g., 30 seconds or more) by adjusting the latent feature dimensions, we observe a noticeable performance degradation when samples significantly exceed the 10-second training window. To improve the model’s generalization across varying audio lengths, incorporating more diverse datasets with variable durations for training and evaluation remains an important aspect of future work. 
\subsubsection{The Number of References} The current model architecture limits the number of reference concept pairs to three, restricting the system’s flexibility when handling extensive or diverse customization. Ablation studies show that \textit{DreamAudio} can be easily fine-tuned to handle a varying number of reference concepts. However, the performance of the system can be affected by the availability of appropriate training data. 
\subsubsection{Artificial Training Data} Due to the lack of real-world data for customized tasks, the audio generated by the proposed model can sometimes sound unnatural. In addition, in many cases, the reference audio naturally corresponds to a subset of the target audio events, which can make the task appear closer to stitching or inpainting in terms of the sound generated. Addressing this limitation requires the development of more diverse and high-quality datasets tailored to specific tasks, allowing models to capture detailed patterns and improve their generalization to practical applications. %Nevertheless, the construction of the customized caption with multiple audio events sometimes gives non-grammatical results, which may confuse the model and reduce the capability of the system on understanding the semantic meaning of the input texts.

\section{Conclusion and Future Works}
% \noindent
In this paper, we have introduced \textit{DreamAudio}, a model for customised text-to-audio generation~(CTTA). Given target prompts and reference audio-caption pairs as input, \textit{DreamAudio} demonstrates satisfying performance in generating audio clips with specified audio content in the CTTA task. In addition, DreamAudio delivers competitive results in the general TTA task, establishing a robust foundation for a wide range of audio-related applications. To further validate the real-world applicability of our method, we created the first benchmark for CTTA, which includes zero-shot audio events to simulate real-world scenarios. Our evaluation demonstrates that \textit{DreamAudio} captures these user-specific audio events, highlighting its ability to generate meaningful and semantically coherent content. In addition, we studied the adaptability of the proposed method, which can effectively handle a varying number of audio-caption pairs as reference concepts. Future work will focus on improving the flexibility of incorporating reference audio-text pairs, changing the number of references, and developing diverse and high-quality real-world data. We will integrate additional modalities such as images and video, and extend its use in other customized audio tasks, such as audio separation, style transfer, and editing.

\section*{Acknowledgments}
\noindent
The authors wish to thank the associate editor and the reviewers for their helpful comments to further improve this work. For the purpose of open access, the authors have applied a Creative Commons Attribution~(CC BY) license to any Author Accepted Manuscript version arising. This research was supported in part by the research scholarship from the China Scholarship Council~(CSC), in part by the internship from ByteDance, in part by British Broadcasting Corporation Research and Development~(BBC R\&D), in part by the Research England ``Games and Innovation Nexus'' programme, in part by the Engineering and Physical Sciences Research Council (EPSRC) under the grants 1EP/T019751/1 and EP/Y028805/1, and in part by a PhD scholarship from the Centre for Vision, Speech and Signal Processing~(CVSSP), University of Surrey. 

% \bibliography{reference}
% \bibliographystyle{IEEEtran}
\bibliographystyle{IEEEtran}
\bibliography{reference}

\begin{IEEEbiography}[{\includegraphics[width=1in,height=1.25in,clip,keepaspectratio]{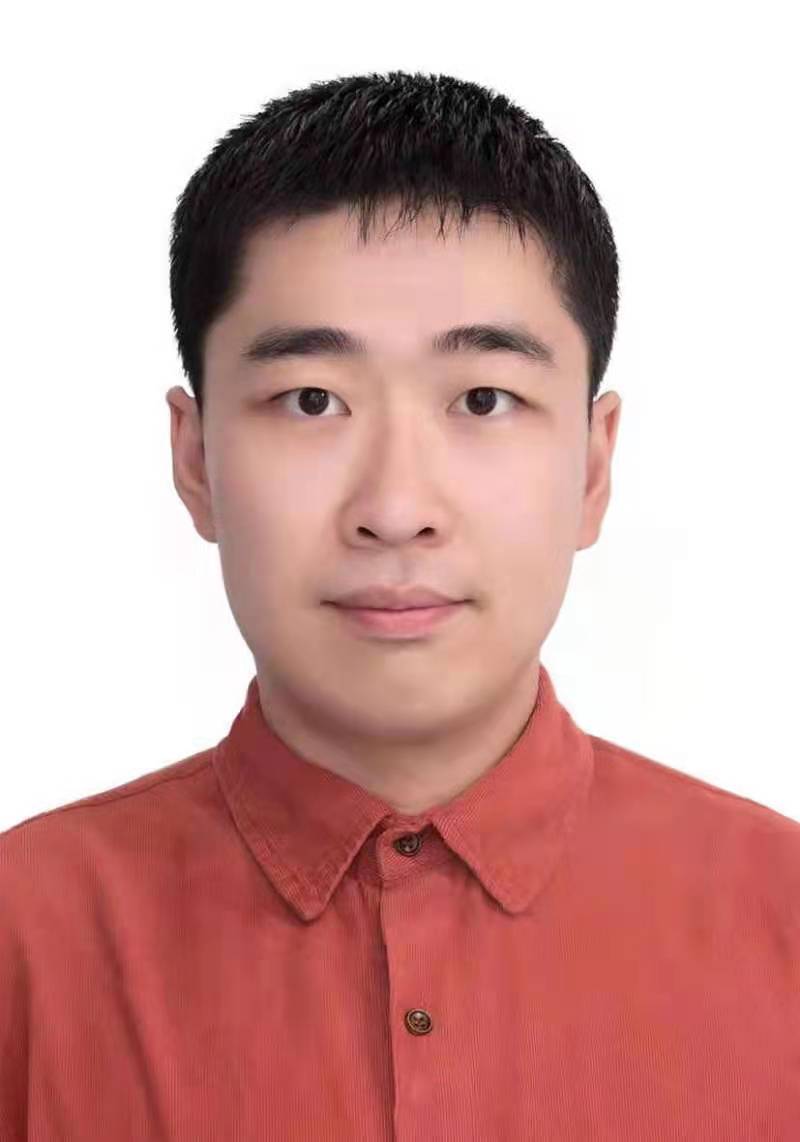}}]{Yi Yuan} received the B.Eng. degree from the University of Sydney, Sydney, NSW, Australia, in 2021, and the M.S. degree in artificial intelligence from the University of Surrey, Guildford, U.K., in 2022. He is currently working toward the Ph.D. degree in vision, speech, and signal processing with the University of Surrey. His research focuses on deep-learning-based audio generation. In 2023, he achieved the top-1 ranking in DCASE Challenge Task 7. He has coauthored more than 20 papers published in leading journals and conferences, including IEEE/ACM Transactions on Audio Speech and Language Processing, IEEE Journal of Selected Topics in Signal Processing, CVPR, ICML, ICASSP, and INTERSPEECH.
\end{IEEEbiography}

\begin{IEEEbiography}[{\includegraphics[width=1in,height=1.25in,clip,keepaspectratio]{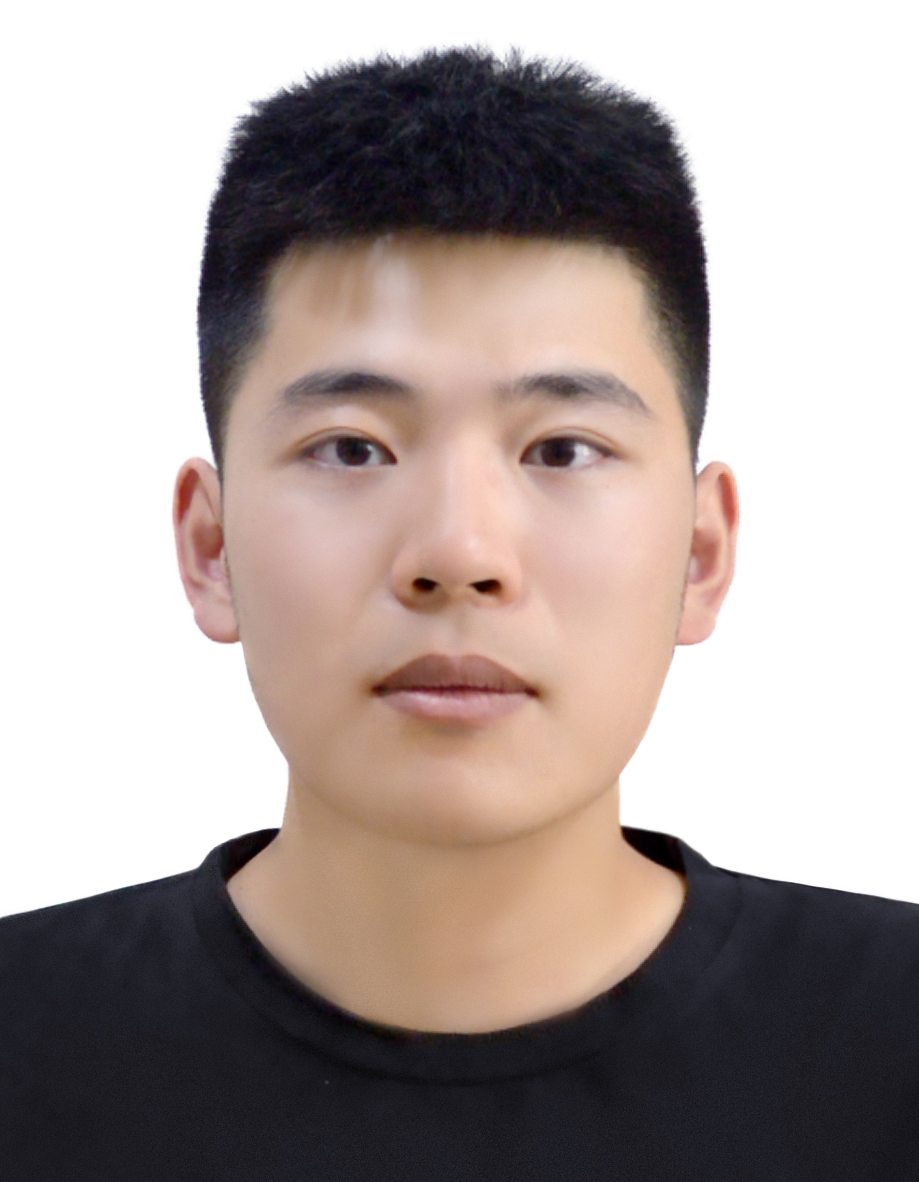}}]{Xubo Liu} received Ph.D.
degree with the Centre for Vision, Speech, and Signal
Processing (CVSSP), University of Surrey, Guildford, U.K. in 2025, working on multimodal learning for audio
and language, focusing on the understanding, separation, and generation of audio signals in tandem with natural language. He has coauthored more than 40 papers in top conferences such as CVPR, ICML, AAAI, EMNLP, ICASSP, and Interspeech. He organized the “Multimodal Learning for Audio and Language” special session at EUSIPCO 2023, and the “Language-Queried Audio Source Separation” challenge on DCASE 2024. He is also a frequent Reviewer for IEEE/ACM Transactions on Audio Speech and Language Processing, CVPR, EMNLP, ICASSP, Interspeech, and MLSP.
\end{IEEEbiography}

\begin{IEEEbiography}[{\includegraphics[width=1in,height=1.25in,clip,keepaspectratio]{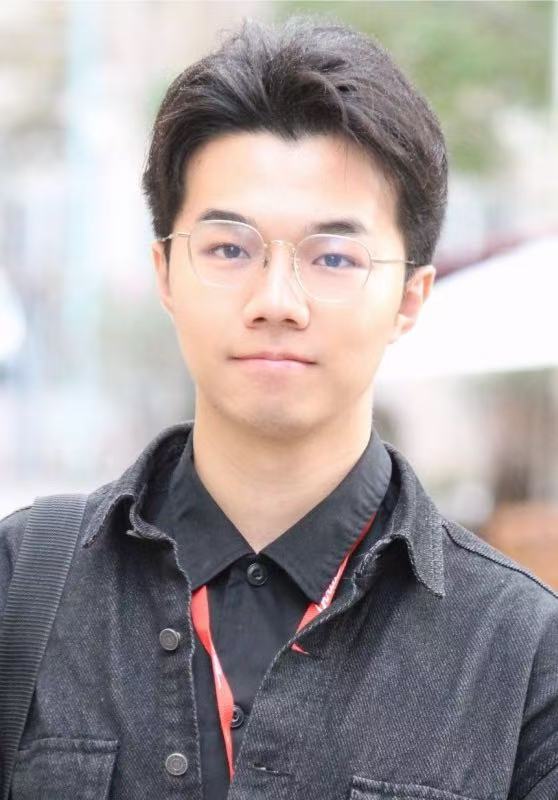}}]{Haohe Liu}  received the B.Eng. degree from Northwestern Polytechnical University, Xi’an, China, in 2020, and the Ph.D. degree 
with the Centre for Vision Speech and Signal Processing, University of Surrey, Guildford, U.K., in 2025. His research has contributed to the fields of audio quality enhancement, audio generation, source separation, and audio recognition. He is best known for developing AudioLDM for text-to-audio generation, which has attracted wide attention in the open-source community. His first-author work has been published in leading journals and conferences such as IEEE Transactions on Pattern Analysis and Machine Intelligence, IEEE/ACM Transactions on Audio, Speech, and Language Processing, IEEE Journal of Selected Topics in Signal Processing, ICML, AAAI, ICASSP, and INTERSPEECH. Notable projects include AudioLDM, VoiceFixer, AudioSR, and NaturalSpeech.
\end{IEEEbiography}

\begin{IEEEbiography}[{\includegraphics[width=1in,height=1.25in,clip,keepaspectratio]{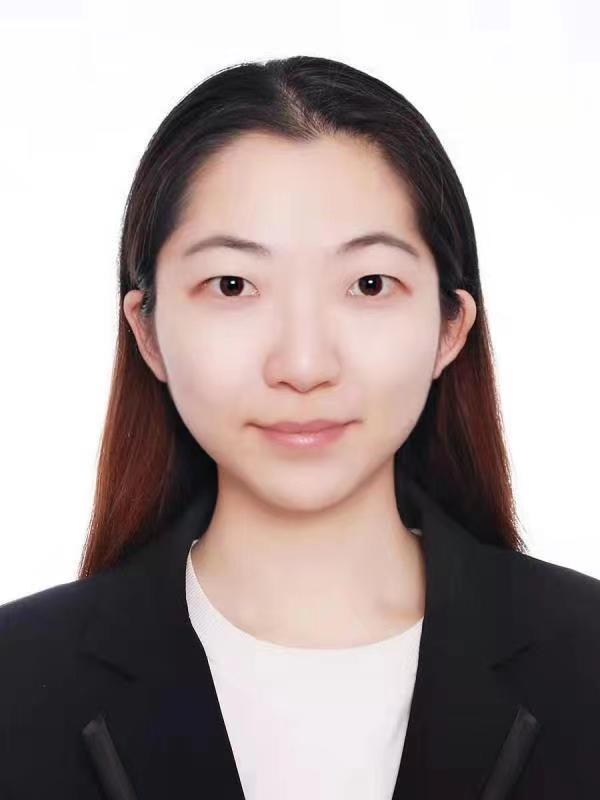}}]{Xiyuan Kang} received the B.Eng. degree from Harbin Engineering University, Harbin, China, in 2021, and the M.S. degree in artificial intelligence from the University of Surrey, Guildford, U.K., in 2023. She is currently pursuing the Ph.D. degree in computer science with the University of Surrey. Her research interests include 3D human pose estimation and shape disentanglement. Her work has been published in leading international conferences, including CVPR. She serves as a reviewer for IEEE Transactions on Neural Networks and Learning Systems.
\end{IEEEbiography}

\begin{IEEEbiographynophoto}{Zhuo Chen}
biography not available at the time of publication.
\end{IEEEbiographynophoto}

\begin{IEEEbiographynophoto}{Yuxuan Wang}
biography not available at the time of publication.
\end{IEEEbiographynophoto}

\begin{IEEEbiography}[{\includegraphics[width=1in,height=1.25in,clip,keepaspectratio]{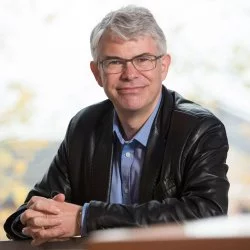}}]{Mark D. Plumbley} (S'88-M'90-SM'12-F'15) received the B.A.(Hons.) degree in electrical sciences and the Ph.D. degree in neural networks from University of Cambridge, Cambridge, U.K., in 1984 and 1991, respectively. He is a Professor of Signal Processing and Head of Department of Informatics department at King's College London, UK. His current research concerns AI, machine learning and signal processing for analysis, recognition and generation of sound. He led the first international data challenge on Detection and Classification of Acoustic Scenes and Events (DCASE) and recently held an Engineering and Physical Sciences Research Council (EPSRC) Fellowship on "AI for Sound". He currently co-leads the EPSRC-funded Noise Network Plus, and is part of the EPSRC AI Hub in Generative Models. He is a Member of the IEEE Signal Processing Society Technical Committee on Audio and Acoustic Signal Processing, and a Fellow of the IET and IEEE.
\end{IEEEbiography}

\begin{IEEEbiography}[{\includegraphics[width=1in,height=1.25in,clip,keepaspectratio]{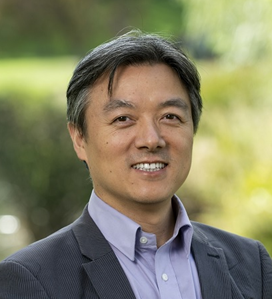}}]{Wenwu Wang} (M’02-SM’11-F’26) was born in Anhui, China. He received the B.Sc., M.E., and the Ph.D. degrees, all in the field of automation, from Harbin Engineering University, China, in 1997, 2000, and 2002, respectively. He then worked with King’s College London, Cardiff University, Tao Group Ltd. (now Antix Labs Ltd.), and Creative Labs, before joining University of Surrey, U.K., in May 2007, where he is currently a Professor in Signal Processing and Machine Learning, and an Associate Head in External Engagement, School of Computer Science and Electronic Engineering, University of Surrey, UK. He is also a Principal AI Fellow at the Surrey Institute for People Centred Artificial Intelligence. His current research interests include signal processing, machine learning and perception, artificial intelligence, machine audition (listening), human-AI collaboration, and statistical anomaly detection. He has (co)-authored over 400 papers in these areas. His works have been recognized with various awards, including the Meta Distinguished Faculty Award (2026), Audio Engineering Society Best Technical Paper Award (2025), IEEE Signal Processing Society Young Author Best Paper Award (2022), DCASE Judge’s Award (2020, 2023, and 2024), DCASE Reproducible System Award (2019 and 2020), and LVA/ICA Best Student Paper Award (2018). He has been elected to IEEE Fellow for contributions to audio classification, generation and source separation, since 2026. He is a Senior Area Editor (2025-2027) for IEEE Open Journal of Signal Processing and an Associate Editor (2024-2028) for IEEE Transactions on Multimedia. He was a Senior Area Editor (2019-2023) and Associate Editor (2014-2018) for IEEE Transactions on Signal Processing, and an Associate Editor (2020-2025) for IEEE/ACM Transactions on Audio Speech and Language Processing. He was the elected Chair (2023-2024) of IEEE Signal Processing Society (SPS) Machine Learning for Signal Processing (MLSP) Technical Committee, and a Board Member (2023-2024) of IEEE SPS Technical Directions Board. He is currently the elected Chair (2025-2027) of the EURASIP Technical Area Committee on Acoustic Speech and Music Signal Processing, a Technical Directions Board Representative (2026-2028) and Executive Sub-Committee Member (2026) of the IEEE SPS Conference Board, and an elected Member (2021-2026) of the IEEE SPS Signal Processing Theory and Methods Technical Committee. He was on the organization committee of IEEE ICASSP 2019 and 2024, INTERSPEECH 2022, IEEE MLSP 2013 and 2024, and IEEE SSP 2009. He was a Technical Program Co-Chair of IEEE MLSP 2025. He has been a keynote or plenary speaker at about 30 international conferences and workshops.
\end{IEEEbiography}

\vfill

\end{document}